\documentclass[12pt]{article}
\usepackage{amsmath}
\usepackage{amssymb}
\usepackage{graphicx}
\usepackage{subfigure}
\usepackage{color}
\usepackage{enumerate}
\usepackage{caption}
\usepackage{natbib}
\usepackage{url} % not crucial - just used below for the URL
\usepackage{hyperref} % allow you to click and jump to the place
\usepackage{float}
\usepackage{amsthm} % nice proof setup
\usepackage{multirow}
\usepackage{algorithm} % pseudo code
\usepackage{algpseudocode}

\definecolor{Pink}{rgb}{1.0, 0.5, 0.5}
\definecolor{Maroon}{rgb}{0.8, 0.0, 0.0}

\def\boxit#1{\vbox{\hrule\hbox{\vrule\kern6pt\vbox{\kern6pt#1\kern6pt}\kern6pt\vrule}\hrule}}

\def\0{{\bf 0}}
\def\A{{\bf A}}

\def\B{{\bf B}}

\def\C{{\bf C}}

\def\D{{\bf D}}

\def\E{{\bf E}}

\def\I{{\bf I}}

\def\L{{\bf L}}

\def\P{{\bf P}}
\def\Q{{\bf Q}}
\def\U{{\bf U}}
\def\V{{\bf V}}
\def\R{{\bf R}}
\def\S{{\bf S}}

\def\w{{\bf w}}
\def\X{{\bf X}}
\def\J{{\bf J}}
\def\x{{\bf x}}

\def\Y{{\bf Y}}
\def\y{{\bf y}}
\def\Z{{\bf Z}}
\def\z{{\bf z}}

\newcommand{\bone}{\mbox{\bf 1}}

\newcommand{\bDelta}{\mbox{\boldmath $\Delta$}}

\newcommand{\bGamma}{\mbox{\boldmath $\Gamma$}}

\newcommand{\bmu}{\mbox{\boldmath $\mu$}}
\newcommand{\bSig}{\mbox{\boldmath $\Sigma$}}

\newcommand{\diag}{\mathrm{diag}}

\def\beqn{\begin{eqnarray}}
\def\eeqn{\end{eqnarray}}
\def\beqns{\begin{eqnarray*}}
\def\eeqns{\end{eqnarray*}}

\def\0{{\bf 0}}
\def\A{{\bf A}}

\def\C{{\bf C}}

\def\D{{\bf D}}

\def\E{{\bf E}}

\def\I{{\bf I}}

\def\R{{\bf R}}

\def\U{{\bf U}}
\def\V{{\bf V}}
\def\S{{\bf S}}

\def\w{{\bf w}}
\def\X{{\bf X}}
\def\x{{\bf x}}

\def\Y{{\bf Y}}
\def\y{{\bf y}}
\def\Z{{\bf Z}}
\def\z{{\bf z}}
\def\1{{\bf 1}}
\def\Q{{\bf Q}}
\def\J{{\bf J}}

\def\trans{^{\rm T}}
\def\strans{^{*\rm T}}
\def\ntrans{^{n\rm T}}
\newtheorem{thm}{Theorem}[section]
\newcommand{\beq}{\begin{equation}}
\newcommand{\eeq}{\end{equation}}
\newcommand{\bes}{\begin{eqnarray*}}
\newcommand{\ees}{\end{eqnarray*}}
\newcommand{\bi}{\begin{itemize}}
\newcommand{\ei}{\end{itemize}}

\newcommand{\bSigma}{\boldsymbol{\Sigma}}
\newcommand{\bLambda}{\boldsymbol{\Lambda}}

\usepackage{tikz}
\usetikzlibrary{shapes}
\usetikzlibrary{positioning,arrows}
\usepackage[utf8x]{inputenc}
\usepackage{scalefnt}
\usetikzlibrary{shapes,shapes.geometric,arrows,fit,calc,positioning,automata}
\usetikzlibrary{shapes,arrows,shadows}

\newcommand{\blind}{1}
\usepackage{authblk}
\usepackage[margin=1in]{geometry}

\usepackage{tikz}
\usetikzlibrary{shapes}
\usetikzlibrary{positioning,arrows}
\usepackage[utf8x]{inputenc}
\usepackage{scalefnt}
\usetikzlibrary{shapes,shapes.geometric,arrows,fit,calc,positioning,automata}
\usetikzlibrary{shapes,arrows,shadows}

\begin{document}

\def\spacingset#1{\renewcommand{\baselinestretch}%
{#1}\small\normalsize} \spacingset{1}

%\pagenumbering{roman}
%%%%%%%%%%%%%%%%%%%%%%%%%%%%%%%%%%%%%%%%%%%%%%%%%%%%%%%%%%%%%%%%%%%%%%%%%%%%%%

\if1\blind
{
\title{Multivariate Log-Contrast Regression with Sub-Compositional Predictors: Testing the Association Between Preterm Infants' Gut Microbiome and Neurobehavioral Outcomes}

\author[1]{Xiaokang Liu}
\author[2]{Xiaomei Cong}
\author[3]{Gen Li}
\author[4]{Kendra Maas}
\author[1]{Kun Chen\thanks{Corresponding author. \href{mailto:kun.chen@uconn.edu}{kun.chen@uconn.edu}.}}
\affil[1]{Department of Statistics, University of Connecticut, Storrs, CT}
\affil[2]{School of Nursing, University of Connecticut, Storrs, CT}
\affil[3]{Department of Biostatistics, Columbia University}
\affil[4]{Microbial Analysis, Resources, and Services, University of Connecticut, Storrs, CT}
%\affil[3]{Center for Population Health, University of Connecticut Health Center, Farmington, CT}
\date{}
  \maketitle
} \fi

\if0\blind
{
  \bigskip
  \bigskip
  \bigskip
  \begin{center}
    {\LARGE Multivariate Log-Contrast Regression with Sub-Compositional Predictors: Testing the Association Between Preterm Infants' Gut Microbiome and Neurobehavioral Outcomes}
\end{center}
  \medskip
} \fi

\bigskip
\begin{abstract}
The so-called gut-brain axis has stimulated extensive research on microbiomes. One focus is to assess the association between certain clinical outcomes and the relative abundances of gut microbes, which can be presented as sub-compositional data in conformity with the taxonomic hierarchy of bacteria. Motivated by a study for identifying the microbes in the gut microbiome of preterm infants that impact their later neurobehavioral outcomes, we formulate a constrained integrative multi-view regression, where the neurobehavioral scores form multivariate response, the sub-compositional microbiome data form multi-view feature matrices, and a set of linear constraints on their corresponding sub-coefficient matrices ensures the conformity to the simplex geometry. To enable joint selection and inference of sub-compositions/views, we assume all the sub-coefficient matrices are possibly of low-rank, i.e., the outcomes are associated with the microbiome through different sets of latent sub-compositional factors from different taxa.
We propose a scaled composite nuclear norm penalization approach for model estimation and develop a hypothesis testing procedure through de-biasing to assess the significance of different views. Simulation studies confirm the effectiveness of the proposed procedure. In the preterm infant study, the identified microbes are mostly consistent with existing studies and biological understandings. Our approach supports that stressful early life experiences imprint gut microbiome through the regulation of the gut-brain axis.\\

%for exact and nearly view-specific low-rank models 

%Multi-view data are routinely collected in various scientific fields.  We propose an {\em integrative multi-view reduced-rank regression} (iRRR) model to study the relationship between multi-view predictors and multivariate responses. We exploit a composite nuclear norm penalization approach, and devise an alternating direction method of multipliers algorithm for model fitting. Extensions to non-Gaussian, incomplete responses and shrinkage estimation are discussed. Theoretically, we derive non-symptomatic oracle bounds of the estimator. Our results seamlessly bridge group-sparse and low-rank methods, and achieve substantially faster convergence rate. Simulation studies and an application to the Longitudinal Study of Aging further showcase the efficacy of iRRR.\\

\noindent%
\vfill
\end{abstract}

%{\color{blue} Gen, thanks for your great effort. I revised Sections 1 and 2 so far (still needs a lot work). Some materials are extracted from my 2015 NSF proposal. In this version, I hope to emphasize the multi-view setup and the connections between our method, group lasso, and nuclear norm penalization. Let's have a discussion.}

%\newpage
\spacingset{1.5} % DON'T change the spacing!

%%%%%%%%%%%%%%%%%%%%%%%%%%%%%%%%%%%%%%%%%%%%%%%
%%
%% Start From Here
%%
%%%%%%%%%%%%%%%%%%%%%%%%%%%%%%%%%%%%%%%%%%%%%%%
\pagenumbering{arabic}

\section{Introduction}

In recent years, there has been a dramatic increase in survival among preterm infants from 15\% to over 90\% \citep{Fanaroff2003, Stoll2010} due to the advancement in neonatal care. However, studies showed that stressful early life experience, as exemplified by the accumulated stress and insults that the preterm infants encounter during their stay in neonatal intensive care units (NICU), could cause long-term adverse consequences for their neurodevelopmental and health outcomes, e.g., \citet{Mwaniki2012} reported that close to 40\% of NICU survivors had at least one neurodevelopmental deficit that may be attributed to stress/pain at NICU, caused by maternal separations, painful procedures, clustered care, among others. As such, understanding the linkage between the stress/pain and the onset of the altered neuro-immune progress holds the key to reduce the costly health consequences of prematurity. This is permitted by the existence of the functional association between the central nervous system and gastrointestinal tract \citep{carabotti2015gut}. With the regulation of this ``gut-brain axis'', accumulated stress imprints on the gut microbiome compositions \citep{Dinan2012, cong2016focus}, and thus the link between neonatal insults and neurological disorders can be approached through examining the association between the preterm infants' gut microbiome compositions and their later neurodevelopment measurements.

%We use data from a study conducted at a Level IV NICU in the northeast region of the U.S. in which both gut microbiome data and neurodevelopment measurements of preterm infants were collected. 

To investigate the aforementioned problem, a preterm infant study was conducted in a NICU in the northeastern region of the U.S. Stable preterm infants were recruited, and fecal samples were collected during the infant's first month of postnatal age on a daily basis when available \citep{cong2016focus,cong2017influence,sun2018log}. From each fecal sample, bacterial DNA was isolated and extracted, and gut microbiome data were then obtained through DNA sequencing and data processing. Gender, delivery type, birth weight, feeding type, gestational age and and postnatal age were recorded for each infant. Infant neurobehavioral outcomes were measured when the infant reached 36--38 weeks of gestational age, using the NICU Network Neurobehavioral Scale (NNNS). More details on the study and the data are provided in Section \ref{sec:app}.

With the collected data, the assessment of which microbes are associated with the neurobehavioral development of the preterm infants can be conducted through a statistical regression analysis, with the NNNS scores being the outcomes and the gut microbiome compositions as the predictors. There are several unique challenges in this problem. First, the NNNS consists of 13 sub-scales on various aspects of neurobehavioral development, including habituation, attention, and quality of movement. As such, an overall assessment about whether the neurobehavioral development is impacted at all by the gut microbiome calls for a multivariate estimation and testing procedure that can utilize all the sub-scale scores simultaneously. Indeed, our preliminary analysis shows that these sub-scale scores are distinct yet interrelated. A multivariate procedure could result in more accurate estimation and more powerful tests than its univariate counterparts. Moreover, the candidate predictors constructed from the microbiome data are structurally very rich and complex: they are high-dimensional, compositional, and hierarchical. A compositional observation is a multivariate vector with elements being proportions, which are non-negative and satisfy the constraint that their summation is unity. In our problem, the data on bacterial taxa are presented as groups of sub-compositions in conformity with the taxonomic hierarchy of bacteria, i.e., each taxon is represented by a group of compositions at a lower taxonomic rank. These unique features call for a tailored dimension reduction approach that can allow high-dimensional inference to be made at the group level for testing each taxon component.

Compositional data analysis is of great importance in a broad range of scientific fields, including microbiology, ecology and geology. The simplex and non-Euclidean structure of the data impedes the application of many classical statistical methods. Much foundational work on the treatment of compositional data was done by John Aitchison \citep{Aitchison1982}; see \citet{aitchison2003statistical} for a thorough survey. In the regression realm, a foundational work is the linear log-contrast model \citep{aitchison1984log}; in its symmetric form, the response is regressed on the logarithmic transformed compositional predictors and a zero-sum constraint is imposed on the coefficient vector to keep the simplex geometry. %The linear restriction not only avoids the thorny reference selection but also fulfills the sub-compositional coherence rule \citep{aitchison2003statistical}. %Based on isometric log-ratio transformed predictors, \citet{hron2012linear} also considered a linear regression model with compositional explanatory variables in low-dimensional situation.  
Compositional data on microbiome are often high dimensional, as it is common that a sample could produce hundreds of operational taxonomic units \citep{Lin2014}. Various sparsity-inducing penalized estimation methods were proposed to enable the selection of a smaller set of relevant compositions; see, e.g., \citet{Lin2014, shi2016regression, wang2017structured, sun2018log,combettes2019regression}.
%\citet{Lin2014} studied a sparse linear log-constrast model by formulating the problem as a constrained lasso regression \citep{tib1996}.
%With longitudinal compositional observations, \citet{sun2018log} considered a constrained group lasso penalized regression. 
%For detecting potential outliers, \citet{mishra2019robust} further incorporated a mean shift component and proposed a robust sparse linear log-contrast model. 
%In order to provide a rigorous false discovery rate control for the selected covariates, \citet{srinivasan2019compositional} proposed a two-step compositional knockoff filter focused on the model considered in \citet{lin2014variable}.
\citet{shi2016regression} extended the sparse regression model to perform high-dimensional sub-compositional analysis, in which the predictors form several compositional groups according to the taxonomic hierarchy of the microbes; a de-biased estimation procedure was adopted to perform statistical inference. %obtain an asymptotically unbiased estimator of the regression coefficients and its asymptotic distribution for making inference. 
%\citet{wang2017structured} applied a newly proposed tree-structured penalty in a regularized regression to conduct sub-composition selection.  
Another kind of regression methods conducts sufficient dimension reduction or low-rank estimation \citep{tomassi2019sufficient, wang2019sliced}. %For example, \citet{tomassi2019sufficient} developed a sufficient dimension reduction method to find the linear combinations of predictors that contain sufficient information from covariates in predicting the response. \citet{wang2019sliced} achieved dimension reduction through the sliced inverse regression procedure. %See \citet{LiH2015} for a recent comprehensive review on microbiome compositional data analysis.
For unsupervised learning of compositional data, various versions of principal component analysis (PCA) \citep{aitchison1983principal, filzmoser2009principal, scealy2015robust} and factor analysis \citep{filzmoser2009robust} have been developed. We refer to \citet{LiH2015} for a recent comprehensive review on microbiome compositional data analysis. To the best of our knowledge, multivariate regression method and inference procedures on studying the association between multiple outcomes and high-dimensional sub-compositional predictors are still lacking.

To assess the association between the neurobehavioral outcomes of the preterm infants and their gut microbiome compositions during NICU stay, we propose multivariate log-contrast regression with grouped sub-compositional predictors. Motivated by \citet{Lin2014} and \citet{li2018integrative}, we formulate the problem as a constrained integrative multi-view regression, in which the neurobehavioral outcomes form the response matrix, the log-transformed sub-compositional data form the multi-view feature matrices, and a set of linear constraints on their corresponding coefficient matrices ensure the obedience of the simplex geometry of the compositions. The linear constraints are then conveniently absorbed through parameter transformation.
To enable joint sub-compositional dimension reduction and selection, we assume that the sub-coefficient matrices are possibly of low ranks. This assumption induces a parsimonious and highly interpretable model for dealing with high-dimensional grouped sub-compositions, i.e., the outcomes are associated with the microbes through different sets of latent sub-compositional factors from different bacterial taxa, and a taxon becomes irrelevant to the outcomes when its corresponding sub-coefficient matrix is a zero matrix or equivalently of zero rank. We develop a scaled composite nuclear norm penalization approach for model estimation and a high-dimensional hypothesis testing procedure through a de-biasing technique. We stress that the proposed approach is generally applicable for a wide range of multivariate multi-view regression problems, and to the best of our knowledge, our work is among the first to develop statistical inference methods for testing high-dimensional low-rank coefficient matrices.

The rest of the paper is organized as follows. Section \ref{sec:model} proposes the multivariate log-contrast model, where the implication of the integrative low-rank structure on analyzing sub-compositional predictors is elaborated. Section \ref{sec:estimation} develops a scaled composite nuclear norm penalization approach for estimating both the mean structure and the variance. Efficient computational algorithms and theoretical guarantees on the resulting estimators are presented. Section \ref{sec:inf} develops the inference procedure and its related theoretical results. Simulation study of the proposed inference procedure is shown in Section \ref{sec:sim}. Section \ref{sec:app} details the application in the preterm infant study. A few concluding remarks and future research directions are provided in Section \ref{sec:fut}. 

\section{Multivariate Log-Contrast Model with Sub-Compositional  Predictors}\label{sec:multilogmodel}

%Standard regression models typically require transformation of the data first, which may lead to reduced power and hampered interpretation.
\subsection{Model}\label{sec:model}

%so were in this preterm infant study

Our work was motivated by the need of identifying gut microbiome taxa during the early postnatal period of preterm infants that may impact their later neurobehavioral outcomes. Microbiome data commonly manifest themselves as compositions. Concretely, a $p$ dimensional compositional vector represents the relative abundances of $p$ different taxa in a sample, and its entries are %strictly positive
non-negative and sum up to one. Therefore, the data are multivariate in nature and reside in a simplex that does not admit the familiar Euclidean geometry. In regression analysis with compositional covariates, the log-ratio transformations are commonly adopted to lift the compositions from the simplex to the Euclidean space, which assumes the data lie in a strictly positive simplex $\z_i\in \mathbb{S}^{p-1}=\{[z_{i1},\ldots,z_{ip}]\trans\in \mathbb{R}^p;~z_{ij}>0,\sum_{j=1}^p z_{ij}=1\}$. In practice, preprocessing steps such as replacing zero counts with some small numbers (e.g., the maximum rounding error) are applied. In this work we adopt this pragmatic log-contrast regression approach in our work. %, with a remark that there exist more sophisticated model-based approaches to deal with zero observations which is a future direction to extend our work \citep{xia2013logistic, LiH2015}. 

Another important feature of microbiome data is the presence of the evolutionary history charted through a taxonomic hierarchy. The major taxonomic ranks are domain, kingdom, phylum, class, order, family, genus and species, from the highest to the lowest. Such a structure provides crucial information about the relationship between different microbes and proves useful in various analyses \citep{shi2016regression, sun2018log}. In practice, selecting the taxonomic rank or ranks at which to perform the statistical analysis depends on both the scientific problem of interest itself and the tradeoff between data quality and data resolution: the lower the rank, the higher the resolution of the taxonomic categories, but the sparser the data for each category. A good compromise is achieved by the sub-compositional regression analysis \citep{shi2016regression}, in which the effect of a taxon on the outcome at the rank of primary interest is investigated through its more information-rich sub-compositions at a lower taxonomic rank. %by using a more delicate sub-compositional structure at a higher taxonomic rank.

In the preterm infant study, the microbiome data can be presented as sub-compositional data of different bacterial taxa at the order level, each consists of a group of compositions at the genus level. To formulate, suppose we have $K$ taxa, and within the $k$-th taxon there are $p_k$ many taxa that are of a lower rank. Let $z_{k,i,j}$ be the subcomposition of the $j$-th genus under the $k$-th order for the $i$-th observation, $\z_{k,i} = [z_{k,i,1},\ldots, z_{k,i,p_k}]\trans\in \mathbb{R}^{p_k}$ be the compositional vector of the $k$-th order for the $i$-th observation, and $\Z_k=[\z_{k,1},\ldots,\z_{k,n}]\trans\in \mathbb{R}^{n \times p_k}$ be the data matrix of the $k$-th order. As such, the integrated sub-compositional design matrix, i.e., $\Z = [\Z_1,\ldots,\Z_K]$, naturally admits a grouped or multi-view structure, and it satisfies that
$
\z_{k,i}\in \mathbb{S}^{p_k-1}, \ k=1,\ldots,K; i=1,\ldots,n.
$
Let $\widetilde\Z_k=\log(\Z_k)$ and $\widetilde\Z = \log(\Z)$ be the corresponding log-transformed sub-compositional data, where $\log(\cdot)$ is applied entrywisely. Also let $\Z_0 \in \mathbb{R}^{n\times p_0}$ be the data matrix of control variables, e.g., gender and birth weight.

In this work, we concern multivariate outcomes, e.g., the 13 sub-scale NNNS scores in the preterm infant study. Let $\Y \in \mathbb{R}^{n \times q}$ be the response matrix consisting of data collected from the same $n$ subjects on $q$ outcome variables. 
We now propose the multivariate log-contrast model with grouped sub-compositional predictors,
\begin{align}\label{sub:eq1}
	\Y= \1_n\bmu\strans + \Z_0\C_0^* + \sum_{k=1}^K \widetilde\Z_k \C^*_k + \E, \qquad ~\mbox{s.t.} \ \bone_{p_k}\trans \C_k^* =\0,~k=1,\ldots,K,
\end{align}
where $\bmu^*\in\mathbb{R}^q$ is the intercept vector, %$\widetilde\Z_k$ is the log-transformed compositional matrix of the $k$-th taxon, and $\C_k^*$ is the corresponding coefficient sub-matrix.
$\C_k^*$ is the $k$-th coefficient sub-matrix for each $k=0,\ldots, K$ and $\E$ is the random error matrix with zero mean and standard deviation $\sigma$. The linear constraints are to ensure that the model obeys the simplex geometry. In what follows we consider two different ways to express the model in an unconstrained form. The first method is to write the model in terms of log-ratio transformed compositional predictors. Specifically, consider taking the first component from each sub-composition as the baseline taxon and eliminating the constraint by using $\C_{k,1}^*=-\sum_{j=2}^{p_k} \C_{k,j}^*$ with $\C_{k,j}^*$ being the $j$-th row in $\C_k^*$. Then an equivalent form of (\ref{sub:eq1}) in terms of log-ratios is
\begin{align}\label{sub:eq01}
	\Y=\1_n\bmu\strans + \Z_0\C_0^* + \sum_{k=1}^K \widetilde\Z_k^1 \C^*_{k \setminus 1} + \E,
\end{align}
where $\widetilde\Z_k^1=(\log( z_{k,i,j}/z_{k,i,1}) )$ is an $n \times (p_k -1)$ log-ratio matrix with the first component as the baseline and $\C_{k \setminus 1}^*=(\C_{k,2}^*,\ldots,\C_{k,p_k}^*)\trans$ is the corresponding $(p_k-1) \times q$ coefficient matrix. As such, the choice of the baseline taxa may lead to inconsistency in model estimation when regularization is adopted \citep{Lin2014}. In contrast, model (\ref{sub:eq1}) can be regarded as a symmetric form of the log-contrast model which avoids any arbitrary selection of baselines.

%Now we consider the estimation of model (\ref{sub:eq1}) under the linear constraints and the taxon-specific low rank assumption of $\C_k^*$, for $k=1,\ldots, K$.

The second way of avoiding the linear constraints is through a linear transformation of the parameters. Let's first rewrite the linear constrains to be 
\begin{align*}
	\L\trans\C=\0, \quad \C=(\C_1\trans,\ldots,\C_K\trans)\trans, \quad  \L=\diag\{ \bone_{p_1}, \ldots, \bone_{p_K} \}
\end{align*}
and write the set of solutions to $\L\trans\C=\0$ as $\{ (\I_{p} - \P_{\mathbb{C}(\L)})\B: \B \in \mathbb{R}^{p \times q} \}$ where $\P_{\mathbb{C}(\L)}$ is the orthogonal projection matrix of the column space of $\L$. Define $\X=\widetilde \Z (\I_{p} - \P_{\mathbb{C}(\L)})=(\X_1,\ldots,\X_K)$, we obtain 
%\begin{align*}
%	\min_{\B \in \mathbb{R}^{p \times q}, \L\trans\B=\0} \|\Y-\Z\B\|_F^2 = \min_{\C \in \mathbb{R}^{p \times q}} \|\Y - \widetilde \Z \C\|_F^2,
%\end{align*}
%i.e., we obtain 
an unrestricted model
\begin{align}\label{sub:nores}
	\Y=\1_n\bmu\strans + \Z_0\C_0^* + \sum_{k=1}^K \X_k \B^*_k + \E, 
\end{align}
where $\X$ is the projected design matrix and $\B^*=(\B_1\strans,\ldots,\B_K\strans)\trans$ is the corresponding coefficient matrix. From the specific form of $\L$, the linear constraints are imposed on each coefficient sub-matrix separately and thus the transformed design matrix $\X=(\X_1,\ldots,\X_K)$ still keeps the original grouping structure of the sub-compositions, so assessing the effect of the $k$-th taxon can be done through testing $\mbox{H}_0: \X_k\B_k^*=\0$. In fact, the transformation on each $\widetilde\Z_k$ is equivalent to doing a centered log-ratio transformation \citep{aitchison2003statistical} to the sub-compositions. %In fact, the transformation on each $\widetilde\Z_k$ is equivalent to doing a simple row-wise centering within each set of sub-compositions. 
Henceforth, we focus on this unrestricted model formulation in (\ref{sub:nores}).

To facilitate dimension reduction and model interpretation, we assume that each $\B_k^*$ is possibly of low rank. %, i.e., $\mbox{rank}(\B_k^*) = r_k^* \leq \min(p_k,q)$. 
That is, model \eqref{sub:nores} exhibits a taxon-specific low-rank structure. %As the covariates naturally form a multi-view structure in the sub-compositional regression model (\ref{sub:eq1}), then by assuming each sub-composition to be of low-rank, we obtain an integrative multivariate log-contrast model.  
%\begin{align}\label{consIRRR}
%\widehat \B^n =\arg \min_{\B \in \mathbb{R}^{p \times q}} \{ \|\Y-\Z\B\|_F^2+\lambda\sum_{k=1}^K w_k\|\B_k\|_*\}, \ s.t. \ \bone_{p_k}'\B_{k}=\0,~k=1,\ldots,K.
%\end{align}
%It is natural to work with a view-specific low-rank structure in sub-compositional data analysis, since both view selection and latent variable extraction are meaningful in the practical problem.
%The relationship between multi-view covariates and the responses can be analyzed in two levels simultaneously under the supervision of responses. One is from the aspect of group selection. Intuitively, some of the predictor sets are influential to the response, while some sets are not. Another is from the aspect of the latent variable extraction from covariates within each view. There might be some correlation between covariates in the same view since they all belong to the same higher rank taxon, and the covariates in a group may affect the responses not in a sparse form but in a collective way. Exacting some latent factors from each predictor set under the supervision of responses can help us avoid information redundancy and control model complexity. Moreover, it is also meaningful to impose low-rank structures on coefficient sub-matrix even when each group is consisting of compositional covariates.
Specifically, suppose the rank of each coefficient sub-matrix is $\mbox{rank}(\B_k^*)=r_k^* \leq \min(p_k,q)$, for $k=1,\ldots,K$. {We can then write $\B_k^*=\J_k \R_k\trans$ as its full-rank decomposition, where $\J_k\in \mathbb{R}^{p_k \times r_k^*}$ and $\R_k\in \mathbb{R}^{q \times r_k^*}$ are both of full column rank. %Due to the sub-compositional nature of the covariates, we have several linear constraints on $\B_k$, i.e., the sum of all rows of $\B_k$ is $\0$. 
%From the relationship $\C_k^*=(\I_{p_k}-\1_{p_k}\1_{p_k}\trans/p_k)\B_k^*$, we have that $\X_k\J_k= \widetilde\Z_k(\I_{p_k}-\1_{p_k}\1_{p_k}\trans/p_k)\J_k$.
%If we write $\A_k$ by its rows $\A_{k,i,\cdot}$ 
%$$
%\A_k=\begin{bmatrix}
% \A_{k,1,\cdot}\trans \\
% \A_{k,2,\cdot}\trans \\
% \vdots \\
% \A_{k,p_k,\cdot}\trans
% \end{bmatrix}
%$$
%then we have
%$$
%\C_k=\begin{bmatrix}
% \A_{k,1,\cdot}\trans \\
% \A_{k,2,\cdot}\trans \\
% \vdots \\
% \A_{k,p_k,\cdot}\trans
% \end{bmatrix} \times \D_k\trans 
% =\begin{bmatrix}
% \A_{k,1,\cdot}\trans \D_k\trans \\
% \A_{k,2,\cdot}\trans \D_k\trans \\
% \vdots \\
% \A_{k,p_k,\cdot}\trans \D_k\trans
% \end{bmatrix}.
%$$
%Thus from $\sum_{j=1}^{p_k} \C_{k,j,\cdot}=\0,~k=1,\ldots,K$, we have 
%$$
%(\A_{k,1,\cdot}\trans+\A_{k,2,\cdot}\trans+\cdots+\A_{k,p_k,\cdot}\trans)\D_k\trans=\0,~k=1,\ldots,K.
%$$
%Recall that $\D_k$ is of full column rank, so we can conclude that 
%$$
%\A_{k,1,\cdot}\trans+\A_{k,2,\cdot}\trans+\cdots+\A_{k,p_k,\cdot}\trans=\0,~k=1,\ldots,K.
%$$
Thus $\X_k^*=\X_k\J_k=\widetilde\Z_k (\I_{p_k}-\1_{p_k}\1_{p_k}\trans/p_k)\J_k$ provides a few latent factors of the original log-transformed data and maintains the compositional structure since it still holds that $\1_{p_k}\trans(\I_{p_k}-\1_{p_k}\1_{p_k}\trans/p_k)\J_k=\0$.} These latent factors 
share the same structure as the principal components constructed in \citet{aitchison1983principal}, where a log linear contrast form of PCA for compositional data was proposed to extract informative compositional proportions; see, also, \citet{aitchison2005compositional}. %, pawlowsky2015modeling}.
However, PCA is unsupervised and utilizes no information from the response, and a naive PCA of all compositional data ignores the sub-compositional structure that embodies the taxonomic hierarchy. Here, the taxon-specific multi-view low-rank structure differs in two aspects, as illustrated in Figure \ref{fig:scaledirrr}. First, the components $\X_k^*$ are jointly predictive of the response since their estimation is under the supervision of $\Y$. Second, the dimension reduction is conducted in a taxon-specific fashion to make use of the structural information and facilitate model interpretation.

\pgfdeclarelayer{background}
\pgfdeclarelayer{foreground}
\pgfsetlayers{background,main,foreground}
% Define block styles used later
\tikzstyle{xu}=[draw, fill=black!20, text width=2.5em, % 2.5em,
    text centered, minimum height=10em,drop shadow]
    
\tikzstyle{xv}=[draw, fill=black!5, text width=6em,
    text centered, minimum height=2.5em,drop shadow]
    
\tikzstyle{xj}=[draw, fill=black!0, text width=2.5em,
    text centered, minimum height=5em,drop shadow]

\tikzstyle{ann} = [above, text width=5em, text centered]

\tikzstyle{by} = [xu, text width=6em, fill=black!20,
    minimum height=10em, rounded corners, drop shadow]
    
\tikzstyle{bx} = [xu, text width=5em, fill=black!10, %text width=6em,
    minimum height=10em, rounded corners, drop shadow]
    
\tikzstyle{myarrows}=[line width=0.5mm,draw=black,-triangle 45,postaction={draw, line width=1.5mm, shorten >=4mm, -}]

\usetikzlibrary{arrows, decorations.markings}
\tikzstyle{vecArrow} = [thick, decoration={markings,mark=at position
   0.8 with {\arrow[semithick]{open triangle 60}}},
   double distance=1.4pt, shorten >= 12pt,
   preaction = {decorate},
   postaction = {draw,line width=1.4pt, white,shorten >= 11pt}]
   
\tikzstyle{innerWhite} = [semithick, white,line width=1.4pt, shorten >= 11pt]

\usetikzlibrary{arrows,positioning}
\tikzset{
    %Define standard arrow tip
    >=stealth',
    %Define style for boxes
    punkt/.style={
           rectangle,
           rounded corners,
           draw=black, very thick,
           text width=6.5em,
           minimum height=2em,
           text centered},
    % Define arrow style
    pil/.style={
           ->,
           thick,
           shorten <=2pt,
           shorten >=2pt,}
}
\def\blockdist{2}
\def\blockuv{3em}
\def\edgedist{2}

\begin{figure}[htp]
\centering
\begin{tikzpicture}[remember picture, scale=0.6, every node/.style={scale=0.4},every block/.style={scale=0.4}]
%\node [label={[label distance=1cm]30:label}] {Node};
    \node at (-1.8,-3) (by) [by]{$\Y$};
    \path (by.east)+(3,0) node (xu1) [xu] {$\X_1^*$};
    \path (xu1.east)+(\blockuv,0) node (xv1) [xv] {$\R_1\trans$};
    \path (xv1.east)+(0.4,-0.2) node (dots1)[ann] {$+$};
    \path (dots1.east)+(0,0) node (xu2) [xu] {$\X_2^*$};
    \path (xu2.east)+(\blockuv,0) node (xv2) [xv] {$\R_2\trans$};
    \path (xv2.east)+(1,-0.2) node (dots2)[ann] {$+\cdots +$};
    \path (dots2.east)+(0.3,0) node (xuk) [xu] {$\X_K^*$};
    \path (xuk.east)+(\blockuv,0) node (xvk) [xv] {$\R_K\trans$};
    
    %\path (xu1.south)+(0,-1.2) node (l1)[ann] {$\1_{p_1}\trans\J_1=\0$};
    %\path (xu2.south)+(0,-1.2) node (l2)[ann] {$\1_{p_2}\trans\J_2=\0$};
    %\path (xuk.south)+(0,-1.2) node (l3)[ann] {$\1_{p_K}\trans\J_K=\0$};

    \path (xu1.north)+(0,3.6) node (x1) [bx] {$\X_1$ };
    \path (x1.east)+(0.8,0) node (xj1) [xj] {$\J_1$};
    \path (xu2.north)+(0,3.6) node (x2) [bx] {$\X_2$ };
    \path (x2.east)+(0.8,0) node (xj2) [xj] {$\J_2$};
    \path (xuk.north)+(0,3.6) node (xk) [bx] {$\X_K$ };
    \path (xk.east)+(0.8,0) node (xjk) [xj] {$\J_K$};
    \path (x2.east)+(2,-0.2) node (dots4) [ann] {$\cdots\cdots$};
    
    \path (x1.north)+(0,2.5) node (z1) [bx] {$\Z_1$ };
    \path (x2.north)+(0,2.5) node (z2) [bx] {$\Z_2$ };
    \path (xk.north)+(0,2.5) node (zk) [bx] {$\Z_k$ };
    \path (dots4.north)+(0,3.5) node (dots5) [ann] {$\cdots\cdots$};
    
    %\path (xu1.south)+(0,-1.2) node (l1)[ann] {$\1_{p_1}\trans\J_1=\0$};
    %\path (xu2.south)+(0,-1.2) node (l2)[ann] {$\1_{p_2}\trans\J_2=\0$};
    %\path (xuk.south)+(0,-1.2) node (l3)[ann] {$\1_{p_K}\trans\J_K=\0$};

    \draw [pil,dashed](z1.south)--(x1.north);
    \draw [pil,dashed](z2.south)--(x2.north);
    \draw [pil,dashed](zk.south)--(xk.north);

\draw [pil,dashed](x1.south)--(xu1.north);
\draw [pil,dashed](x2.south)--(xu2.north);
\draw [pil,dashed](xk.south)--(xuk.north);

\draw [pil,dashed](xj1.south)--(xu1.north);
\draw [pil,dashed](xj2.south)--(xu2.north);
\draw [pil,dashed](xjk.south)--(xuk.north);

\draw [vecArrow](xu1.west)+(-0.6,0)--(by.east);
\draw [innerWhite](xu1.west)+(-0.6,0)--(by.east);

 \draw (by.north)
   edge[pil, black, bend left=15,dashed] (xu1.north) % edges are used to connect two nodes
   edge[pil, black, bend left=15,dashed] (xu2.north)
   edge[pil, black, bend left=15,dashed] (xuk.north); % .east since we want

\node [draw=black, fit= (xu1) (xv1) (xuk) (xvk),inner sep=0.25cm,scale=1/0.4] {};

\end{tikzpicture}
\caption{Diagram of the taxon-specific low-rank multivariate log-contrast model with grouped sub-compositional predictors. Latent taxon-specific features $\X_k^*$ are learned from each log-transformed sub-compositions under the compositional constrains and the supervision of $\Y$.}\label{fig:scaledirrr}
\vspace{-0.35cm}
\end{figure}
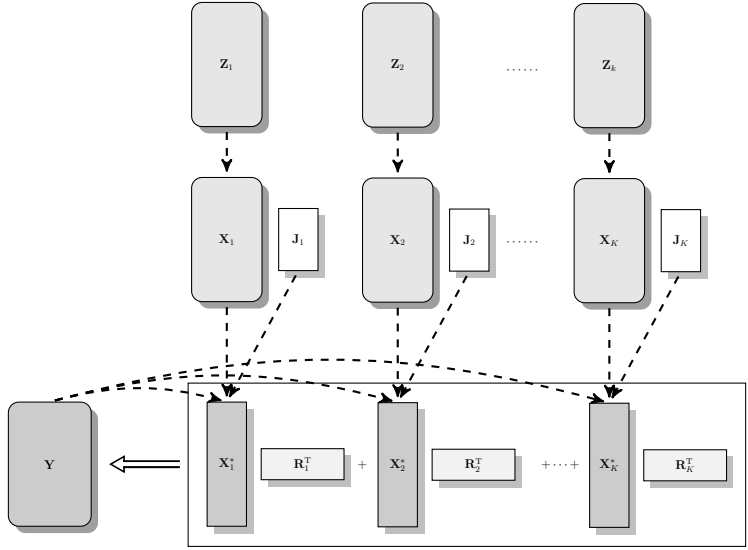

\subsection{Estimation via Scaled Composite Nuclear Norm Penalization}\label{sec:estimation}

Model (\ref{sub:nores}) can be recognized as an integrative reduced-rank regression (iRRR) model proposed by \citet{li2018integrative}, in which a composite nuclear norm penalization approach was developed for estimating the regression coefficients. However, due to the need for enabling statistical inference, the estimation of the error variance and the adaptive estimation of the coefficient matrix are both crucial. Therefore, following the scaled lasso framework \citep{sun2012scaled},
we develop a scaled composite nuclear norm penalization approach,  %the original iRRR estimation framework \citep{li2018integrative}
%\begin{align}
%\widehat \B^n(\w) = \arg \min_{\B \in \mathbb{R}^{p \times q}} \L_{\w}(\B)= \arg \min_{\B \in \mathbb{R}^{p \times q}} \left\{ \frac{1}{2nq} \|\Y - \X\B\|_F^2+\lambda\sum_{k=1}^K w_k \|\B_k\|_* \right\}, \label{eq:main1}	
%\end{align} {\color{red} Please do not repeat the same formula if possible. Besides, why make this comparison even before presenting scaled RRR? }
%and its scaled version
\begin{align}
(\widehat\bmu, \widehat\C_0, \widehat \B^n, \hat \sigma ) %& = \arg \min_{\bmu, \C_0, \B, \sigma} \L_{\w}(\bmu, \C_0, \B, \sigma) 
 = \arg \min_{\bmu, \C_0, \B, \sigma} \left\{ \frac{1}{2nq \sigma} \|\Y - \1_n\bmu\trans - \Z_0\C_0 - \X\B\|_F^2 + \frac{\sigma}{2} +\lambda\sum_{k=1}^K w_k \|\B_k\|_* \right\},\label{eq:main2}	
\end{align}
%We specify $\widehat \B^n(\w)$ as the solution of (\ref{eq:main1}) based on the set of weights $\w=(w_1, \ldots, w_K)$. 
where for each $k=1,\ldots, K$, $\|\B_k\|_*$ denotes the nuclear norm of matrix $\B_k$, $\lambda$ is a tuning parameter to control the amount of regularization (no penalization on $\C_0$). We choose the weights as $w_{k}=d_1(\X_k)\{\sqrt{p_kq} + \sqrt{2\log(K/\epsilon)}\}/(nq)$ for some $0<\epsilon<1$, to achieve desired statistical performance (see Theorem \ref{thm:scaleiRRR} below), where $d_j()$ denotes the $j$-th largest singular value of an enclosed matrix. The application of the composite nuclear norm penalty nicely bridges low-rank models and group sparse models. Specifically, this penalty promotes the sparsity of the singular values of each coefficient sub-matrix and could even make the sub-matrix to be entirely zero. With the incorporation of noise level $\sigma$ into the optimization scheme, the resulting coefficient estimator and noise level estimator are scale-equivariant with respect to $\Y$. We have developed efficient algorithms to solve (\ref{eq:main2}), which are presented in Appendix \ref{web:A1}. The resulting estimator is termed as the scaled iRRR estimator. 

%a block-wise coordinate descent algorithm and an alternating direction method of multipliers \citep[ADMM]{boyd2011distributed} algorithm, and the details are provided in Web Appendix A. The resulting estimator is te

%\subsubsection{Theoretical Results}

We investigate the theoretical properties of the scaled iRRR estimator. For simplicity, we present the analysis with the model without the intercept and the control variables
$$
\Y=\X\B^* + \E = \sum_{k=1}^K \X_k\B_k^* + \E,
$$
since the results can be easily extended to the general model \eqref{sub:nores} with a fixed number of controls. A restricted strong convexity (RSC) condition \citep{negahban2012, negahban2011} is exploited to ensure the convexity of the loss function on a restricted parameter space. %, which includes all the possible solutions to the optimization problem (\ref{eq:main2}) once certain conditions on weights and $\lambda$ are satisfied. 
Specifically, the design matrix $\X$ satisfies the RSC condition over a restricted set $C(r_1,\ldots,r_K;\eta,\delta) \in \mathbb{R}^{p \times q}$ if there exists a constant $\kappa(\X)>0$ such that
\begin{align*}
	\frac{1}{2n}\|\X\Delta\|_F^2 \geq \kappa(\X)\|\Delta\|_F^2,\ \text{for all} \ \Delta \in C(r_1,\ldots,r_K;\eta,\delta).
\end{align*}
Here $r_k$ is the rank imposed on each coefficient sub-matrix and satisfies $1 \leq r_k \leq \min(p_k,q)$, $\eta$ is a positive constant and $\delta$ is a tolerance parameter from RSC condition. For details about the restricted set, refer to \citet{li2018integrative}. Next, we give the main theoretical result of the scaled iRRR estimator. 

\begin{thm}\label{thm:scaleiRRR} Assume that $vec(\E) \sim \mathcal{N}_{nq}(\0, \sigma^2 \I_{nq})$. Let $(\widehat \B^n, \hat \sigma)$ be a solution of optimization problem (\ref{eq:main2}), $\B^*$ be the true coefficient matrix, and $\sigma^*=\|\Y-\X\B^*\|_F/ \sqrt{nq}$ be the oracle noise level. Suppose $\X$ satisfies the RSC condition with $\kappa(\X)>0$ over $C(r_1,\ldots,r_K; \eta,\delta)$. When $w_{k} = d_1(\X_k)w_{*,k}/\sqrt{nq}$ with $w_{*,k}=\sqrt{p_k/n} + \sqrt{2\log(K/\epsilon)/(nq)}$ and $0<\epsilon<1$, if we let $\lambda = (1+\theta) (1+\eta)/\sqrt{[ 1- 16(1+\eta)^2(2+\eta)\sum_{k=1}^K q r_k w_k^2/\{\eta^2 \kappa(\X)\} ]_+} $ for any $\theta>0$, then with probability at least $1-\epsilon$,  
%Define the event 
%	$$\mathbb{E} =  \cap_{k=1}^K \mathbb{A}_k = \cap_{k=1}^K \left\{\frac{d_1(\X_k\trans\E)}{nq \sigma^*/\sqrt{1+\tau_-}}   \leq \frac{\lambda w_k}{1+\eta}\right\}.$$
  we have
  $$\sum_{k=1}^K \lambda w_k \| \widehat\B_k^n - \B_k^* \|_* \preceq \frac{\sigma^{*} q \sum_{k=1}^K r_k \lambda^2  w_k^2}{\sqrt{1-\tau_+}\kappa(\X)}, \qquad
  \|\widehat \B^n - \B^*\|_F^2 \preceq \frac{\sigma^{*2} q^2 \sum_{k=1}^K r_k \lambda^2   w_k^2}{(1-\tau_+)\kappa^2(\X)},
  $$
%\begin{align*}
%		\sum_{k=1}^K w_k \| \widehat\B_k^n - \B_k^* \|_* & \preceq \frac{\sigma^{*} q \sum_{k=1}^K r_k w_k^2}{\sqrt{1-\tau_+}\kappa(\X)},  \\
%		 \|\widehat \B^n - \B^*\|_F^2 & \preceq \frac{\sigma^{*2} q^2 \sum_{k=1}^K r_k  w_k^2}{(1-\tau_+)\kappa^2(\X)} 
%\end{align*} 
and
\begin{align}\label{eq:working1}
	%\max\left\{ \left|1-\frac{\hat\sigma}{\sigma^*}\right|, 
	\frac{1}{\sqrt{nq}}\sum_{k=1}^K \frac{\lambda w_{*,k}}{\sigma} \|\X_k\widehat\B_k^n - \X_k\B_k^*\|_F  %\right\}
	=O_p\left(\sum_{k=1}^K q r_k \lambda^2  w_k^2 \right),
\end{align}
when each $\B_k^*$ is exactly of rank $r_k$ and $\tau_+=8(2+\eta)^2  \sum_{k=1}^K q r_k \lambda^2 w_k^2 /\{\eta^2 \kappa(\X)\}$. In addition, if $\sqrt{nq} \sum_{k=1}^K qr_k \lambda^2 w_k^2/\kappa(\X) \rightarrow 0$, we have 
\begin{align}
 \sqrt{nq} \left(\frac{\hat \sigma}{\sigma}-1\right) \rightarrow \mathcal{N}\left(0, \frac{1}{2}\right). \label{asympnormal}
 \end{align}

\end{thm}

%The proof is relegated to the Web Appendix A. Theorem \ref{thm:scaleiRRR} provides the estimation error bounds for the scaled iRRR estimator under the setting that the multi-view components are exactly of low-rank. The results can be readily extended to include an approximation error term when the components are only approximately low rank. Moreover, (\ref{eq:working1}) provides the rate of the weighted mixed prediction error of the scaled iRRR estimator and establishes the consistency and the asymptotic distribution of the noise level estimator $\hat\sigma$.
%The main difference between Theorem \ref{thm:scaleiRRR} here and Theorem 2 in \citet{li2018integrative} is caused by the incorporation of noise level estimation. Consequently, the specific form of $w_k$'s is derived from different probability inequalities in proofs of two theorems. Compare the results in Theorem \ref{thm:scaleiRRR} with the related results from scaled lasso \citep{sun2012scaled} and scaled group lasso \citep{mitra2016benefit}, instead of the RSC condition, they applied a sign-restricted cone invertibility \citep{ye2010rate} type of regularity condition on the design matrix, which is able to produce $\ell_q$-loss error bound for $q \geq 1$. This type of condition is also verified to be weaker than RSC and thus leads to a sharper error bound. Since the target of this paper is to develop a valid inference procedure but not to compete with other methods in terms of sharper error bounds, RSC is good enough to guarantee the whole procedure to work.

{The proof is relegated to the Appendix \ref{web:B}. Theorem \ref{thm:scaleiRRR} provides the error rates of the scaled iRRR estimator $\widehat\B^n$, and establishes the consistency and the asymptotic distribution of $\hat\sigma$. The incorporation of noise level estimation leads to the major difference between Theorem \ref{thm:scaleiRRR} here and Theorem 2 in \citet{li2018integrative}. In particular, the specific forms of $w_k$'s are derived from different probability inequalities in proofs of two theorems. Moreover, Theorem \ref{thm:scaleiRRR} is able to recover the error rates of both the scaled group lasso estimator \citep{mitra2016benefit} and scaled lasso estimator \citep{sun2012scaled}. With the assumption that $\lambda d_1(\X_k)/\sqrt{n} \asymp 1$ and plug in the exact form of $w_k$'s %into (\ref{eq:working1}), 
we have \begin{align}\label{eq:working}
	%\left|1-\frac{\hat\sigma}{\sigma^*}\right| + 
	%\frac{1}{\sqrt{nq}}\sum_{k=1}^K \frac{w_{*,k}}{\sigma} \|\X_k\widehat\B_k^n - \X_k\B_k^*\|_F
	q \sum_{k=1}^K r_k \lambda^2  w_k^2  \asymp \frac{\sum_{k=1}^K r_k\{p_k q + 2 \log(K/\epsilon)\} }{nq}.
\end{align}
By letting $q=1$, (\ref{eq:working}) reduces to the rate of the scaled group lasso estimator in mixed $\ell_2$ loss under a strong group sparsity condition \citep{huang2010benefit}, which is of the order $\{s+g\log(K/\epsilon)\}/n$ with $g$ the number of predictive groups and $s$ the number of entries contained in these groups. If further we let $K=p$ and $p_k=1$ for all $k$, then the rate becomes $s\sqrt{\log(p/\epsilon)/n}$ with $s$ the cardinality of the active set, which is the rate for the scaled lasso in $\ell_1$ loss.

\section{Hypothesis Testing for Sub-Compositional Inference}\label{sec:inf}

We concern the problem of testing %$\mbox{H}_0: \B^*_k=\0  \ \mbox{vs.}\ \mbox{H}_1: \B^*_k \neq \0$ (or 
$\mbox{H}_0: \X_k\B^*_k=\0\ \mbox{vs.}\ \mbox{H}_1: \X_k\B^*_k \neq \0$ under model (\ref{sub:nores}), from which the test result indicates the significance level of the predictive power of the $k$-th group of covariates on the responses when controlling the effects from other covariates. In the preterm infant gut microbiome study, the application of the proposed test can facilitate the identification of potential biomarkers, e.g., bacterial taxa that relate to later neurological disorders with any given level of confidence. One feature of the problem is that we need to test %whether a large set of parameters equal to zeroes simultaneously, while most available methods focus on single coefficient inference... 
whether a group of covariates is predictive at all to multiple responses, while most available methods focus on inference with a single response. Statistical inference for regularized estimators is undergoing exciting development in recent years. Our approach is built upon the theoretical analysis on scaled iRRR and the works by \citet{mitra2016benefit} and \citet{zhang2014confidence} on the low-dimensional projection estimator (LDPE). See Appendix \ref{web:C} for a brief overview of high-dimensional inference procedures and the LDPE approach in particular. The details of our proposed inference procedure are provided in Appendix \ref{web:D}. In what follows, we summarize the main steps of implementing the proposed method. %, which include (1) the exploitation of LDPE to correct the bias of the scaled iRRR estimator, (2) the construction of a $\chi^2$-type test statistic based on the de-biased estimator, and (3) the estimation of the required score matrix and the theoretical guarantee of the reliability of the test.

%Besides, the scaled iRRR estimator is highly nonlinear due to the regularization imposed on the model thus the related asymptotic distribution is difficult to derive.

%First, we provide the de-biased version of the scaled iRRR estimator. 
Let $\S_k\in \mathbb{R}^{n \times p_k}$ be the score matrix of $\X_k$, a critical tool used in LDPE to correct the bias caused by regularization and only depends on $\X$. %{\color{red}This is a critical tool used in LDPE to correct the bias caused by regularization and will be specified later.} 
Write $\Q_k$ and $\P_{0,k}$ be the orthogonal projection matrices onto the column spaces of $\X_k$ ($\mathbb{C}(\X_k)$) and $\S_k$ ($\mathbb{C}(\S_k)$), respectively, and let $\P_k$ be the projection matrix of $\mathbb{C}(\P_{0,k}\Q_k)$. %Hereafter, we assume 
If $\mbox{rank}(\S_k\trans\X_k)=\mbox{rank}(\X_k)$, which guarantees the effectiveness of the de-biasing procedure, %Based on the scaled iRRR estimator $\widehat \B^n=(\widehat \B_1\ntrans,\ldots,\widehat\B_K\ntrans)\trans$ from (\ref{eq:main2}), the de-biased estimator of $\B_k$ is
%\begin{align}
%	\widehat \B_k=\widehat \B_k^n + (\S_k'\X_k)^+\S_k\trans(\Y-\X\widehat\B^n), \label{ldpe1}
%\end{align}   
%where $(\S_k'\X_k)^+$ is the Moore-Penrose inverse of $\S_k'\X_k$. For the group effect $\X_k\B_k$, the related de-biased estimator is 
%\begin{align}
%	\X_k\widehat \B_k & = \X_k\widehat \B_k^n + (\P_k\Q_k)^+\P_k(\Y-\X\widehat\B^n). \label{ldpe2}
%\end{align}
%Next, based on the de-biased estimator, we introduce a test statistic and derive its asymptotic distribution under the null. The effect of de-biasing in $\widehat \B_k$ and $\X_k\widehat\B_k$ is controlled by the approximation of $\S_k$ to $\X_k^{\perp}$ and the distance between $\widehat\B^n$ and $\B^*$, where $\X_k^{\perp}$ is the best score matrix only available in the `low-dimensional' scenario and is defined as the projection of $\X_k$ onto the orthogonal complement of the column space spanned by $(\X_1,\ldots,\X_{k-1},\X_{k+1},\ldots,\X_K)$. These two factors can be jointly measured by  
%\begin{align}\label{remk}
%	\mbox{Rem}_{k}=\P_k\sum_{j \neq k} (\X_j\widehat \B_j^n-\X_j\B_j^*).
%\end{align}
%Once the magnitude of $\mbox{Rem}_{k}$ is ignorable in the sense that 
%\begin{align}
% 	\sqrt{qr_k'}|\sigma/\hat\sigma-1|+\|\mbox{Rem}_k/\sigma\|_F=o_p(1), \label{inf4}
% \end{align}
% where $\hat\sigma$ is from (\ref{eq:main2}) 
then with $r_k'=\mbox{rank}(\P_k)=\mbox{rank}(\X_k)$ and %we have 
 %\begin{align}
%	\|\P_k\E-Rem_k\|_F^2/\hat\sigma^2 \rightarrow \|\P_k\E/\sigma\|_F^2.
%\end{align}
the assumption on the error matrix that $vec(\E) \sim \mathcal{N}_{nq}(\0, \sigma^2 \I_{n q})$,
we have a test statistic 
\begin{align}
	T_k=\frac{1}{\hat \sigma^2}\left\|\P_k(\Y-\sum_{j \neq k}\X_{j}\widehat \B_{j}^n)\right\|_F^2 \overset{H_0}{\sim} \chi^2_{r_k'q}
\end{align}
asymptotically, where $\widehat\B_j^n$ and $\hat\sigma$ are the scaled iRRR estimator and 
$\P_k$ can be estimated from the penalized regression 
\begin{align}
	\widehat \bGamma_{-k} = \arg\min_{\bGamma_{j, j\neq k}} \left\{ 
	\frac{1}{2n} \|\X_k - \sum_{j \neq k} \X_j \bGamma_j \|_F^2 + \sum_{j \neq k} \frac{\xi w_j^{''}}{\sqrt{n}} \|\X_j \bGamma_j\|_* \right\} \label{inf8}
\end{align}
with $w_j''$ the pre-specified weight and $\xi$ a tuning parameter. We estimate the score matrix through $\S_k = \X_k - \X_{-k} \widehat \bGamma_{-k}$ and $\P_k=\S_k(\S_k\trans\S_k)^{-1}\S_k\trans$. The algorithm to solve \eqref{inf8} is provided in Appendix \ref{web:A2}. For the selection of $\xi$, from the comments of \citet{mitra2016benefit}, in practice we only have to find a $\xi$ to make sure $d_1(\P_k(\I_n-\Q_k))<1$, which implies the key condition $\mbox{rank}(\S_k\trans\X_k)=\mbox{rank}(\X_k)$ in de-biasing and testing. The validity of the proposed test is guaranteed by the following result.
\begin{thm}\label{thm2}
  Let $(\widehat\B^n, \hat\sigma)$ be from solving (\ref{eq:main2}), $\P_k$ from (\ref{inf8}) with $w_j^{''}=w_{*,j}$ and $w_{*,j}=\sqrt{p_j/n} + \sqrt{2\log(K/\epsilon)/(nq)}$, $0<\epsilon<1$. The proposed asymptotic hypothesis testing procedure is valid if 
\begin{align}
    	\frac{r_k'}{n}\rightarrow 0,\ \sum_{j=1}^K \frac{r_j \{ p_j q + 2\log(K/\epsilon) \}}{\sqrt{nq}}\left\{\xi d_{min}(\S_k/\sqrt{n})^{-1}  \right\} \rightarrow 0, \label{cond}
\end{align} 
where $d_{min}(\cdot)$ is the smallest singular value of an enclosed matrix.  
\end{thm}

%Finally, we provide a sufficient condition on sample size to guarantee the proposed inference procedure to be valid. %, i.e., the condition in (\ref{inf4}) holds. 

The proof is in Appendix \ref{web:E}. Theorem \ref{thm2} implies that, once the sample size is large enough compared to the test size $r_k'$ and the model complexity, the bias can be ignored and the asymptotic test can provide us reliable inference results. As such, with a pre-fixed significance level $\alpha$, we reject the null hypothesis if $T_k > \chi^2_{\alpha, r_k'q}$, the $\alpha$-th upper quantile of the $\chi^2_{r_k'q}$ distribution. Again, due to the application of the LDPE technique and the generality of the scaled iRRR framework, the derived test can be specialized to solve lasso and group lasso estimator inference problems.

\section{Simulation}\label{sec:sim}

We conduct simulation studies to investigate the performance of the proposed method in making group inference. To show the power gained by multivariate testing, we also apply scaled group lasso testing procedure \citep{mitra2016benefit} to each response and exploit a union test \citep{roy1953heuristic} principle to combine the results, i.e., $\X_k$ is significantly associated with $\Y$, if it is significantly associated with at least one of the $q$ responses in $\Y$ after the Bonferroni correction is applied to control the familywise Type-I error. Three simulation scenarios are considered: (1) the predictors in $\X$ are generated from multivariate normal distributions; (2) we mimic the structure of the preterm infant data to generate compositional predictors which are then processed to produce $\X$, and (3) we directly use the observed compositional data from the preterm infant study in the simulation through resampling with replacement (the details and results are contained in Appendix \ref{web:F}). The latter two are to investigate the behaviors of the proposed method with realistic microbiome data. 

%In addition, as we discussed in Section \ref{sec:inf}, the proposed procedure is very general in the sense that it can cover the inference procedure for lasso \citep{zhang2014confidence} and group lasso \citep{mitra2016benefit} due to the generality of the iRRR method and the shared LDPE idea. Since the performance of the LDPE in scaled lasso and scaled group lasso situation has been comprehensively investigated by \citet{zhang2014confidence} and \citet{mitra2016benefit}, respectively, the focus of this section is on the iRRR model.  
\subsection{Simulation with Normally Distributed Predictors}

%{\color{red} I am curious what was your motivation of designing the following three simulation scenarios. Lots of parameters change and it is hard to make any solid conclusion from comparing these settings. Next time you may want to think more about what you learned from design of experiment.}

We work on two model settings with different dimensionality and complexity:
\begin{itemize}
	\item[1.] $n=500$, $q=5$, $p=50$, $K=5$, $p_i=10,\ i=1,\ldots,5$, and $r_1^*=2,\ r_i^*=0,\ i=2,\ldots,5$.
	%\item[2.] $n=1000$, $q=10$, $p=200$, $K=10$, $p_i=20,\ i=1,\ldots,10$, and $r_i^*=3,\ i=1,2,3$, $r_4^*=\ldots=r_{10}^*=0$.
	\item[2.] $n=200$, $q=10$, $p=400$, $K=20$, $p_i=20,\ i=1,\ldots,10$, and $r_1^*=1$, $r_i^*=0,\ i=2,\ldots,20$.
\end{itemize}  
% The first setting is the classical scenario where we have $n > p$. %, while in setting 1 the number of groups is small ($K=5$) and in setting 2 we increase the group number to be 10. 
% Setting 2 considers a high-dimensional situation where we let $n < p$. %, and we further increase the group number to be 20. 
The design matrix $\X=(\X_1,\ldots,\X_K)\in \mathbb{R}^{n \times p}$, true coefficient matrix $\B^*=(\B_{1}\strans,\ldots,\B_{K}\strans)\trans\in \mathbb{R}^{p \times q}$ and the corresponding response matrix $\Y\in \mathbb{R}^{n \times q}$ are generated as below:
\begin{itemize}
	\item[1.] We generate $\B_{k}^* \in \mathbb{R}^{p_k \times q}$ of rank $r_{k}^*,\ k=1,\ldots,K$ through full-rank decomposition, i.e., $\B_{k}^* = \J_k \R_k\trans $ where $\J_k \in \mathbb{R}^{p_k \times r_{k}^*}$ and $\R_k \in \mathbb{R}^{q \times r_{k}^*}$, and each entry of both $\J_k$ and $\R_k$ is generated from $\mathcal{N}(0,1)$. Then we scale the coefficient matrix to make its largest entry to be 1.
	\item[2.] Each row of $\X$ is generated independently from a multivariate normal distribution $\mathcal{N}_p(\0,\bSig)$. Two covariance structures are considered, (1) within-group autoregressive, i.e., $\bSig$ is block diagonal with diagonal blocks $\bSig_k=(\rho^{|i-j|}) \in \mathbb{R}^{p_k \times p_k}$, and (2) among-group autoregressive, i.e., $\bSig = (\rho^{|i-j|})$. The correlation strength $\rho_x $ is in $\{0, 0.5\}$.	
	\item[3.] The entries of $\E$ are drawn from $\mathcal{N}(0,\sigma^2)$ and the response matrix $\Y$ is obtained from $\Y=\X\B^*+\E$, where $\sigma^2$ is set to control the signal to noise ratio (SNR), defined as the ratio between the standard deviation of the linear predictor $\sum_{k=1}^K \X_k \B_{k}^*$ and the standard deviation of the random error. We consider $\mbox{SNR} \in \{0.1, 0.2, 0.4\}$.
\end{itemize}

%Also, we standardize each $\X_k$ by $d_1(\X_k)/\sqrt{n}$ to simplify the measurement of relative model complexity with respect to the sample size, which is represented as $ \sum_{k=1}^K r_k(p_k q + 2 \log(K))/\sqrt{nq}$. A small $ \sum_{k=1}^K r_k(p_k q + 2 \log(K))/\sqrt{nq}$ ensures an approximately valid inference procedure. The complexity measurements are 2.06 and 4.61 in setting 1 and setting 3, respectively, while in setting 2 it is 18.41. Together with different levels of SNR and correlation strength, in total we consider 6 situations under each setting.

In each replication, we generate $(\X,\Y)$ and conduct group-wise tests with significance level $0.05$. We use 5-fold cross validation to select $\lambda$ in the scaled iRRR and the scaled group lasso, using the negative log-likelihood as the error measure to take into consideration the noise level estimation. As for the estimation of the score matrix, we use $\xi=1$ in (\ref{inf8}) which is verified to be adequate for satisfying $d_1(\P_k(\I_n - \Q_k)) < 1$ in all the settings. Under each setting, the simulation is repeated 100 times.
%For applying scaled iRRR to obtain the initial coefficient matrix estimate and noise level estimate, we use 5-fold cross validation to select $\lambda$ that minimizes the negative log-likelihood function, which as well takes into consideration the noise level estimation.
% we also consider withinCorr
%During the derivation of the de-biased scaled estimator, the score matrix $\P_k$ plays an important role. It measures the unique information that is carried by the set of covariates that is of our interests to do inference. %Thus, it is necessary to consider the effects of different correlation structure among multiple sets of predictors. 
%Similar to the setting considered in \citet{mitra2016benefit}, we impose a group-wise correlation structure on $\X$, where the covariance matrix is a block diagonal matrix with each block $\bSig_k=(\rho^{|i-j|}) \in \mathbb{R}^{p_k \times p_k}$. %In this situation, each view contains more distinct information than the previous described correlation structure. And there is also another tuning parameter involved in the score matrix estimation. 
%We apply the panelized multivariate linear regression of $\X_k$ on $\X_{-k}$ with penalty $ \xi \sum_{j \neq k} \|\X_j\bGamma_j\|_*$ to obtain $\P_k$. Since we only need to make $d_1(\P_k(\I_n - \Q_k)) < 1$ in practice, we choose $\xi=1$ in \ref{inf8} which is verified to be adequate for satisfying this condition in all the settings. Under each setting, the simulation is repeated 100 times. % CV under estimation, small rate rate
% select \xi
% metric
We compute the mean and standard deviation of $\hat\sigma/\sigma-1$ and $|\hat\sigma/\sigma-1|$, respectively, to measure the performance of the noise level estimation. For assessing the inference procedure, we compute both the false positive rate (FP), the proportion of time the test for an irrelevant group is rejected, and the true positive rate (TP), the proportion of time the test for a relevant group is rejected. 

%we select several views from both predictive and irrelevant views to conduct inference. For setting 1 and setting 3, we record the testing results for the first, second and third view, where the first view is predictive and the other two views are irrelative to the response. For setting 2, we record the testing results for the first five views, where the first three views are predictive and the remaining two views have no contribution to the response. False positive (FP) rate and true positive (TP) rate are reported. %Moreover, for verifying the asymptotic distribution of the test statistic, we provide another measurement which is obtained by first plugging in the true value $\B^*_k$ into $T_k$ to obtain $T_k$ with the estimator $\widehat\B$ in each replication, and then computing the rate of this value is less than or equal to the 95\% quantile of $\chi^2_{\mbox{rank}(\P_k)q}$ among 100 replications. We expect the computed rate is close to 95\%. 
We first examine the asymptotic distributions of both $\sqrt{2nq}(\hat\sigma/\sigma-1)$ and the pivotal statistic $\|\P_k\E-\mbox{Rem}_k\|_F^2/\hat\sigma^2$ (refer to Appendix \ref{web:D}) using Setting 1; in the simulation we fix a randomly generated $\X$ with the within-group correlation setup and generate $\E$ in each replication. Plots (a)-(c) in Figure \ref{fig:qq1} display the normal Q-Q plots of $\sqrt{2nq}(\hat\sigma/\sigma-1)$ under different SNR and $\rho_x$ settings. In each plot, the majority of the points approximately lie on a straight line that is coincident with or parallel to the diagonal line. When the SNR is very low, the empirical and theoretical quantiles match quite well, while when the SNR is stronger, the variance is slightly underestimated. This may be due to the nature of the cross validation. Plots (d)-(f) in Figure \ref{fig:qq1} display the $\chi^2$ Q-Q plots to verify the asymptotic distribution of the pivotal statistic. Indeed it approximately follows a $\chi^2$ distribution with degree of freedom $r_k'q$. The slight parallel discrepancy above the diagonal line is caused by the underestimation of the noise level. See Figure \ref{fig:rho0} for the Q-Q plots with $\rho_x=0$. %{\color{red} Which covariance structure did you use?? Please make sure the setting is clear.} 
\begin{figure}[htp]
\centering 
\subfigure[$\mbox{SNR}=0.1,\rho_x=0.5$]{\includegraphics[width=0.32\textwidth]{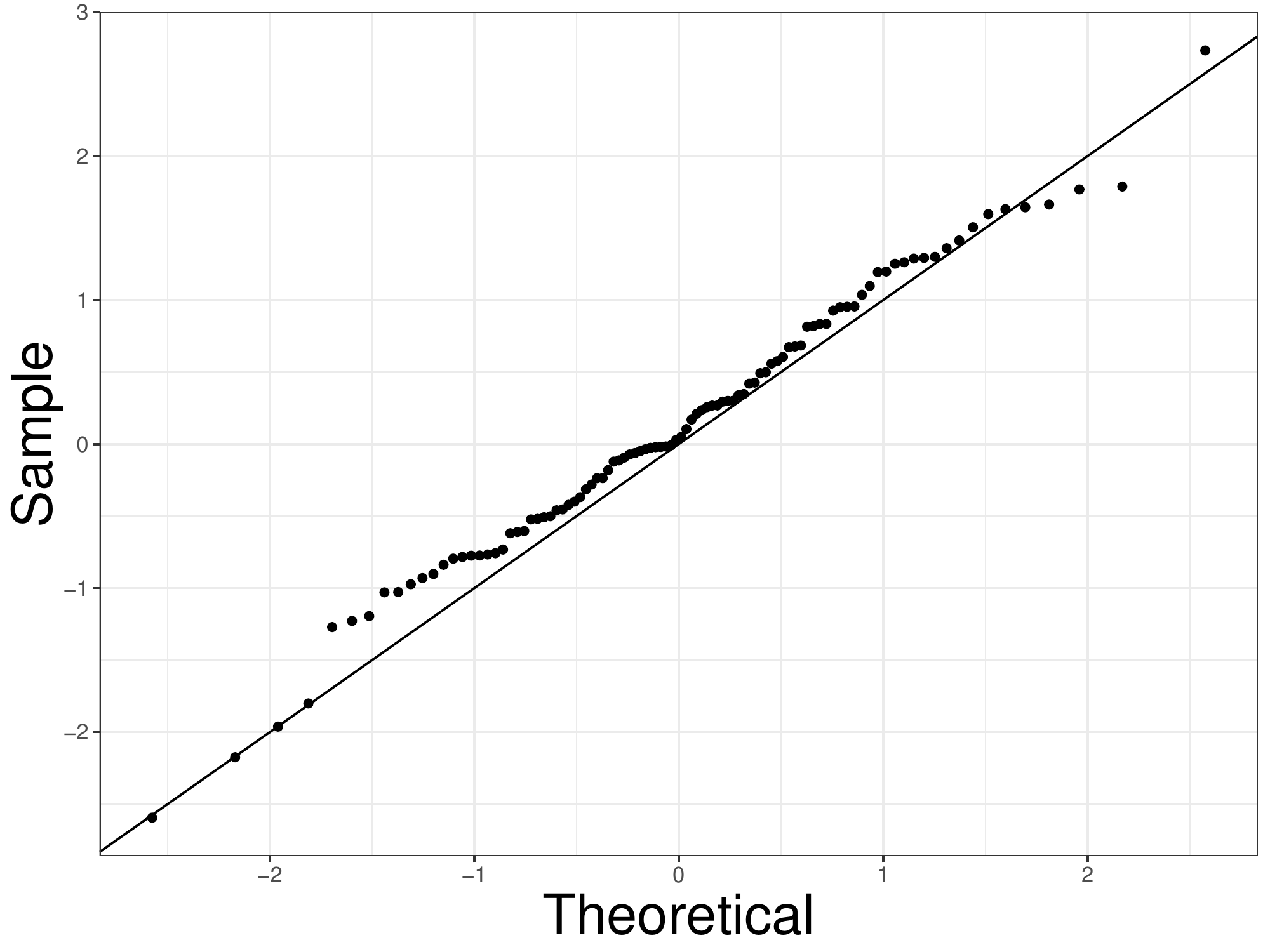}}
\subfigure[$\mbox{SNR}=0.2,\rho_x=0.5$]{\includegraphics[width=0.32\textwidth]{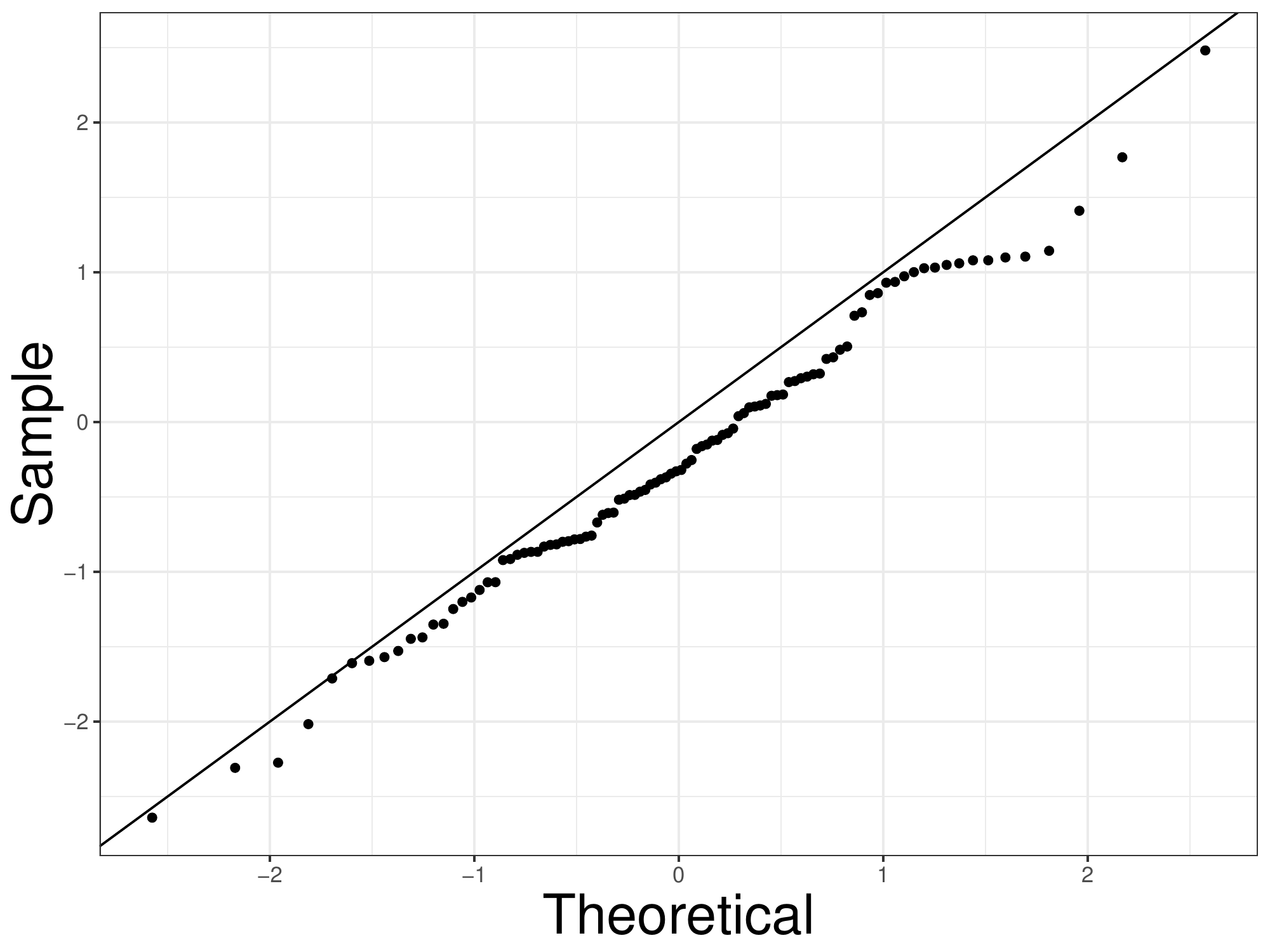}}
\subfigure[$\mbox{SNR}=0.4,\rho_x=0.5$]{\includegraphics[width=0.32\textwidth]{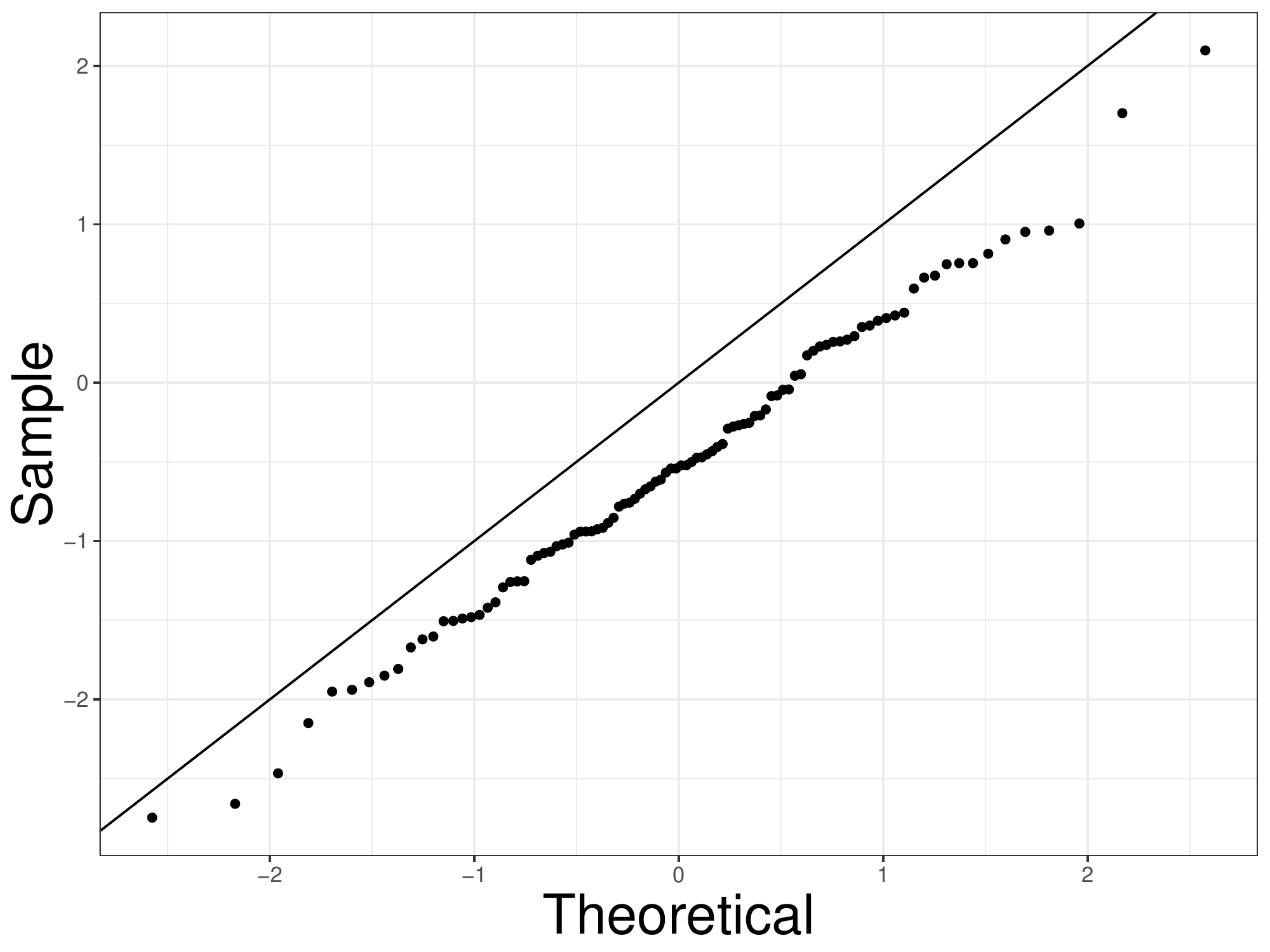}}
\subfigure[$\mbox{SNR}=0.1,\rho_x=0.5,\mbox{group}\ 1$]{\includegraphics[width=0.32\textwidth]{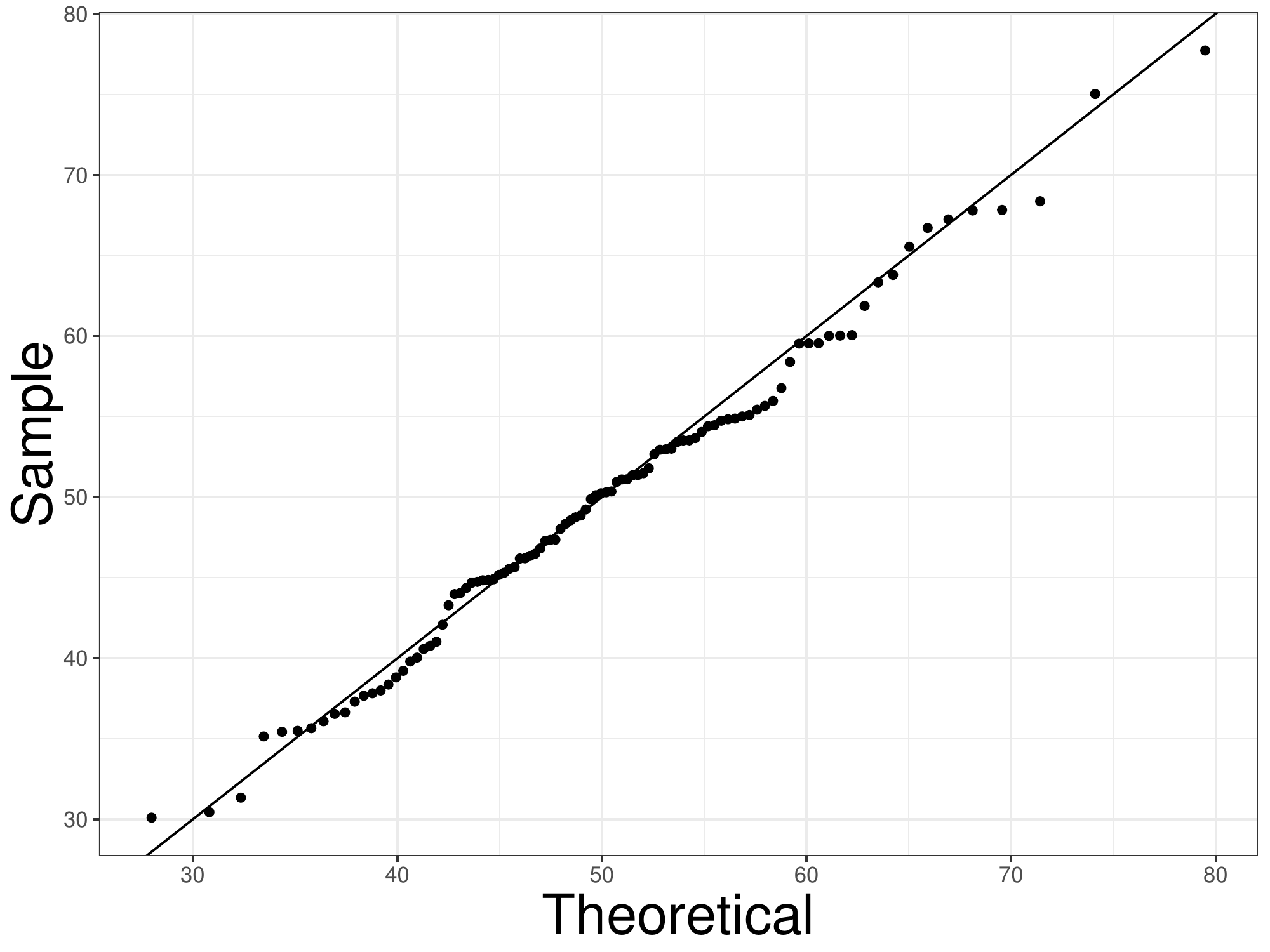}}
\subfigure[$\mbox{SNR}=0.1,\rho_x=0.5,\mbox{group}\ 2$]{\includegraphics[width=0.32\textwidth]{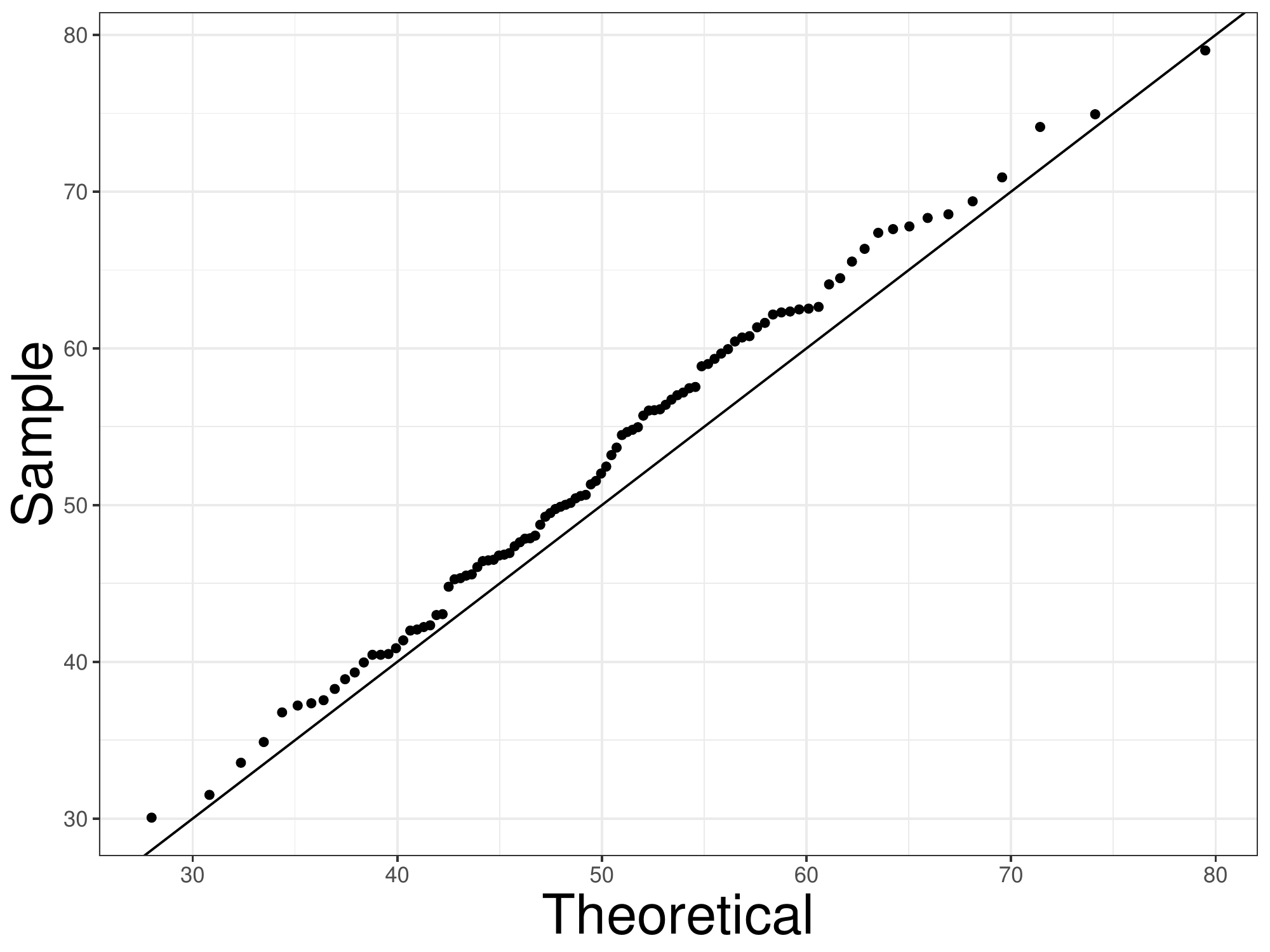}}
\subfigure[$\mbox{SNR}=0.1,\rho_x=0.5,\mbox{group}\ 3$]{\includegraphics[width=0.32\textwidth]{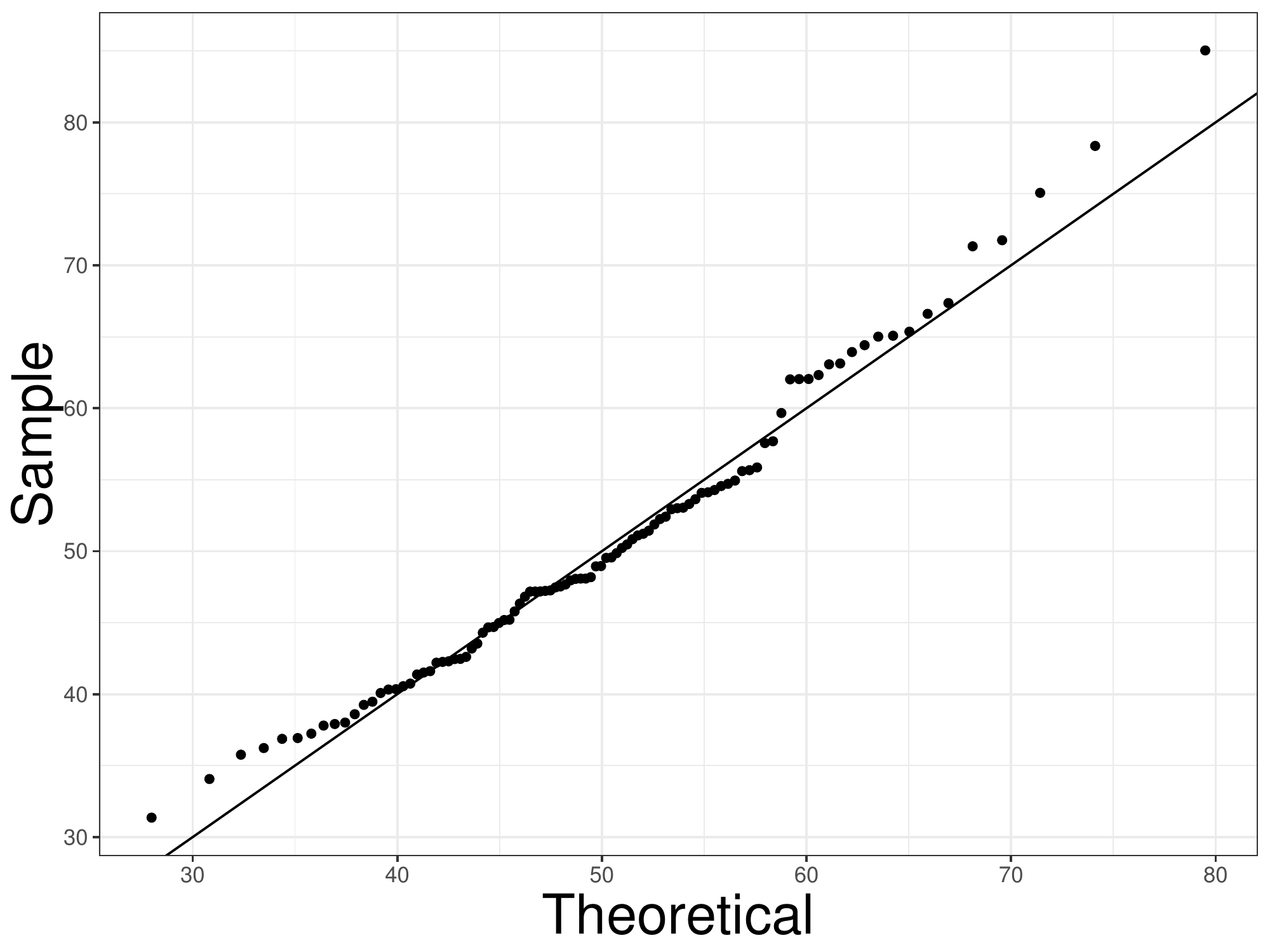}}
\caption{ Simulation results for Setting 1 with $\rho_x=0.5$ from 100 simulation runs: (a) - (c) are the Q-Q plots of $\sqrt{2nq}(\hat\sigma/\sigma-1)$ versus $\mathcal{N}(0,1)$; (d) - (f) are the Q-Q plots of $\|\P_k\E - \mbox{Rem}_k\|_F^2/\hat\sigma^2$ versus $\chi^2_{r_k'q}$.}\label{fig:qq1}
\end{figure}

\begin{figure}[htp]
\centering 
\subfigure[$\mbox{SNR}=0.1,\rho_x=0$]{\includegraphics[width=0.32\textwidth]{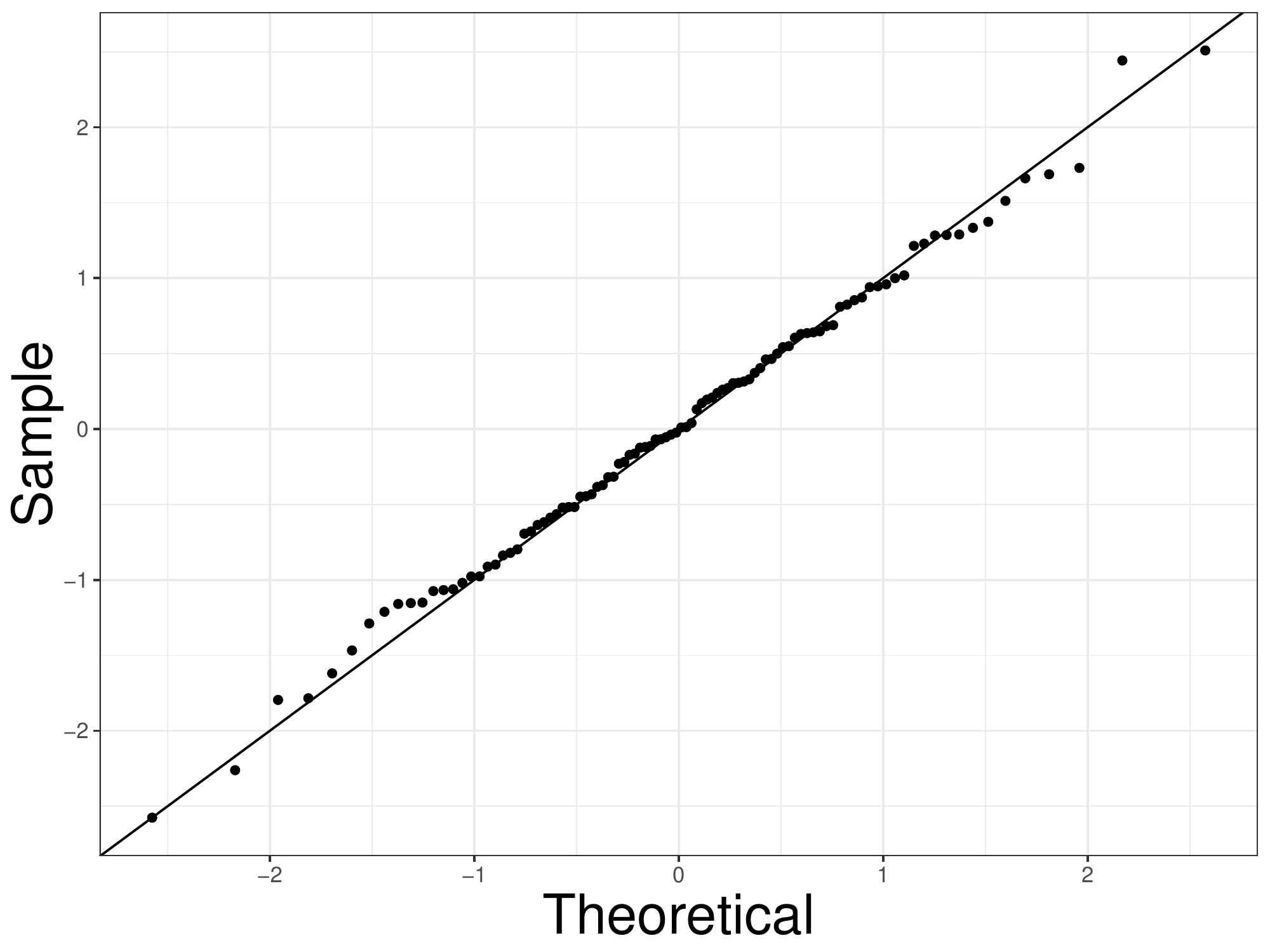}}
\subfigure[$\mbox{SNR}=0.2,\rho_x=0$]{\includegraphics[width=0.32\textwidth]{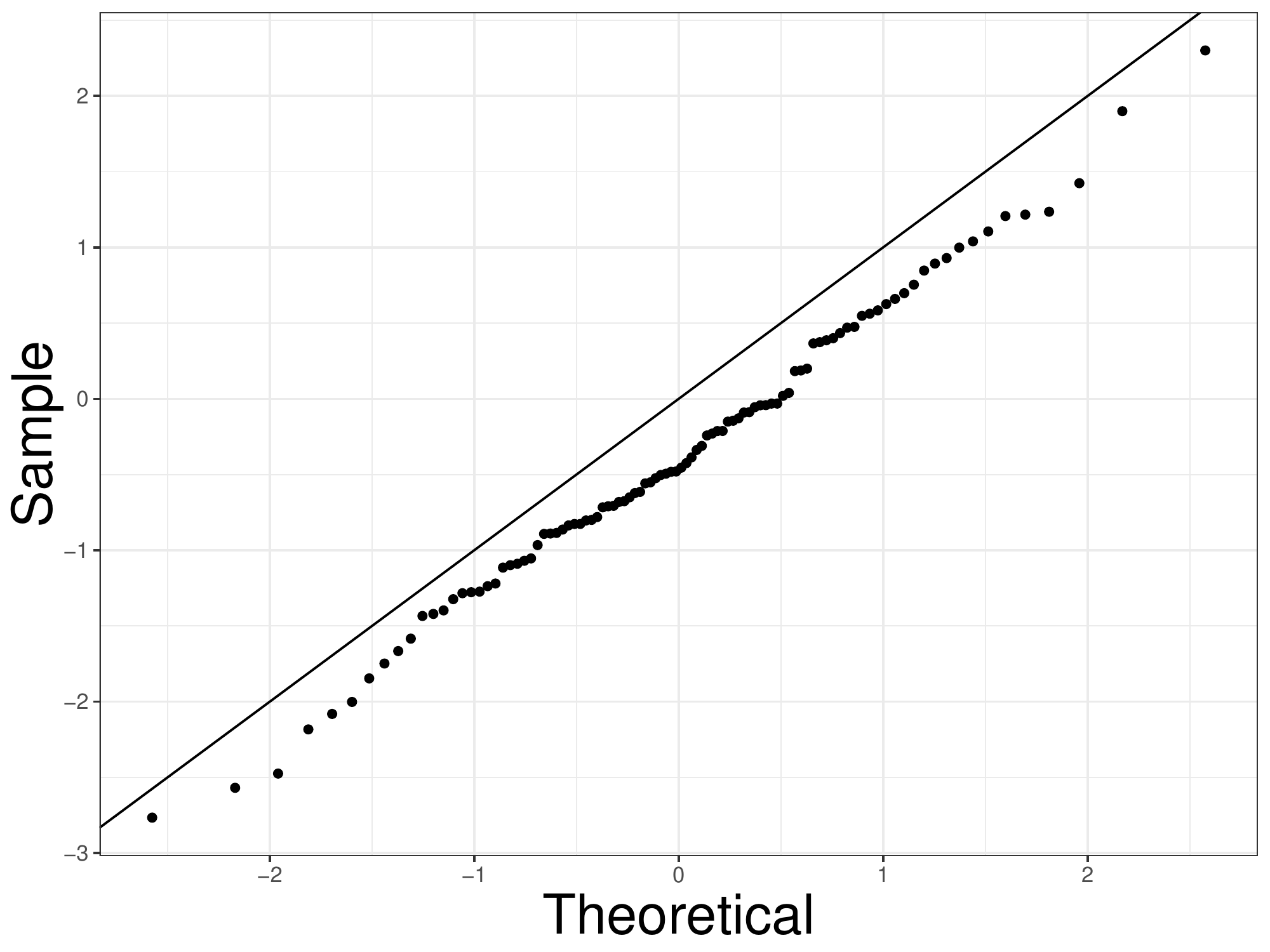}}
\subfigure[$\mbox{SNR}=0.4,\rho_x=0$]{\includegraphics[width=0.32\textwidth]{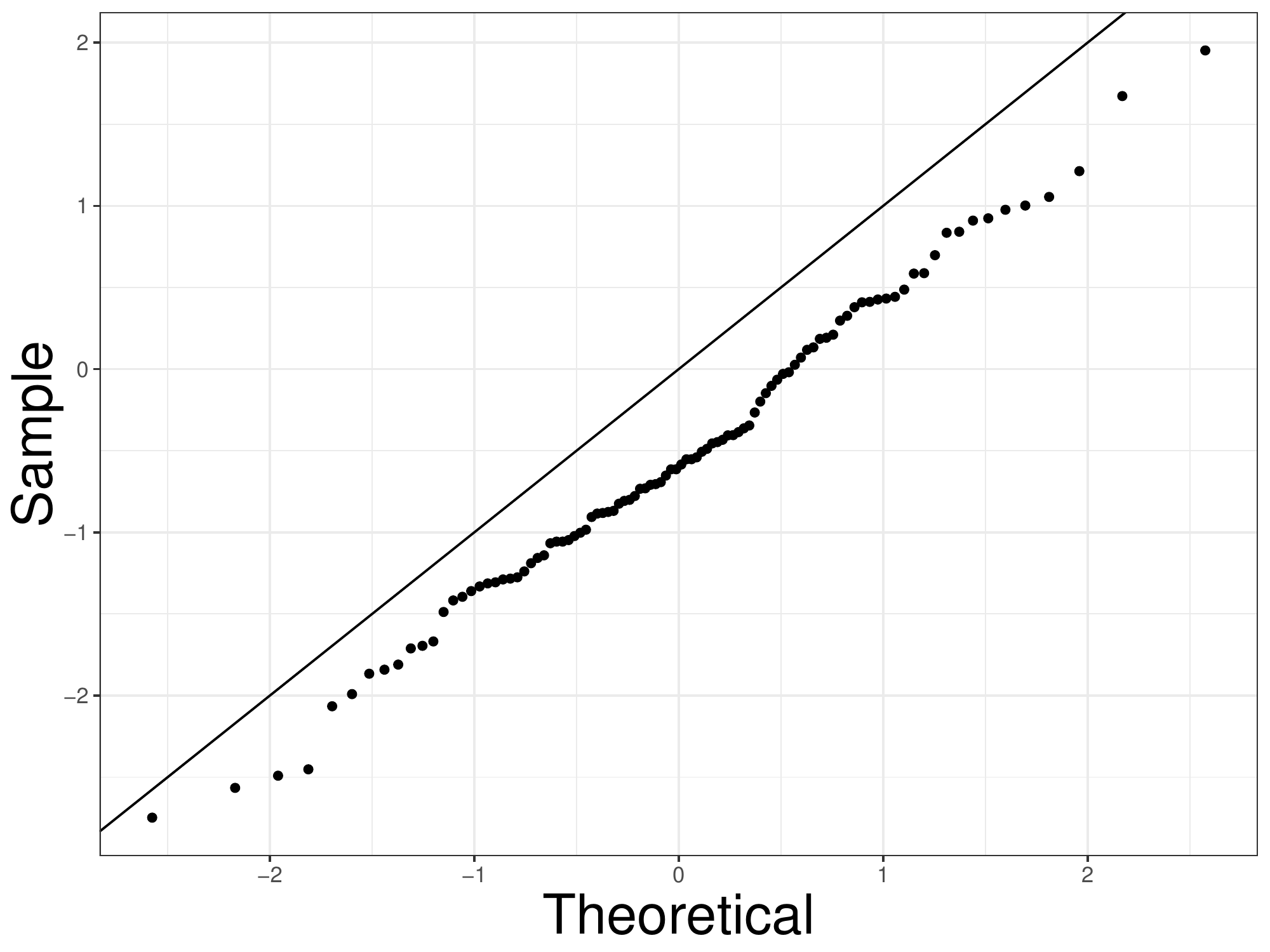}}
\subfigure[$\mbox{SNR}=0.1,\rho_x=0,\mbox{group}\ 1$]{\includegraphics[width=0.32\textwidth]{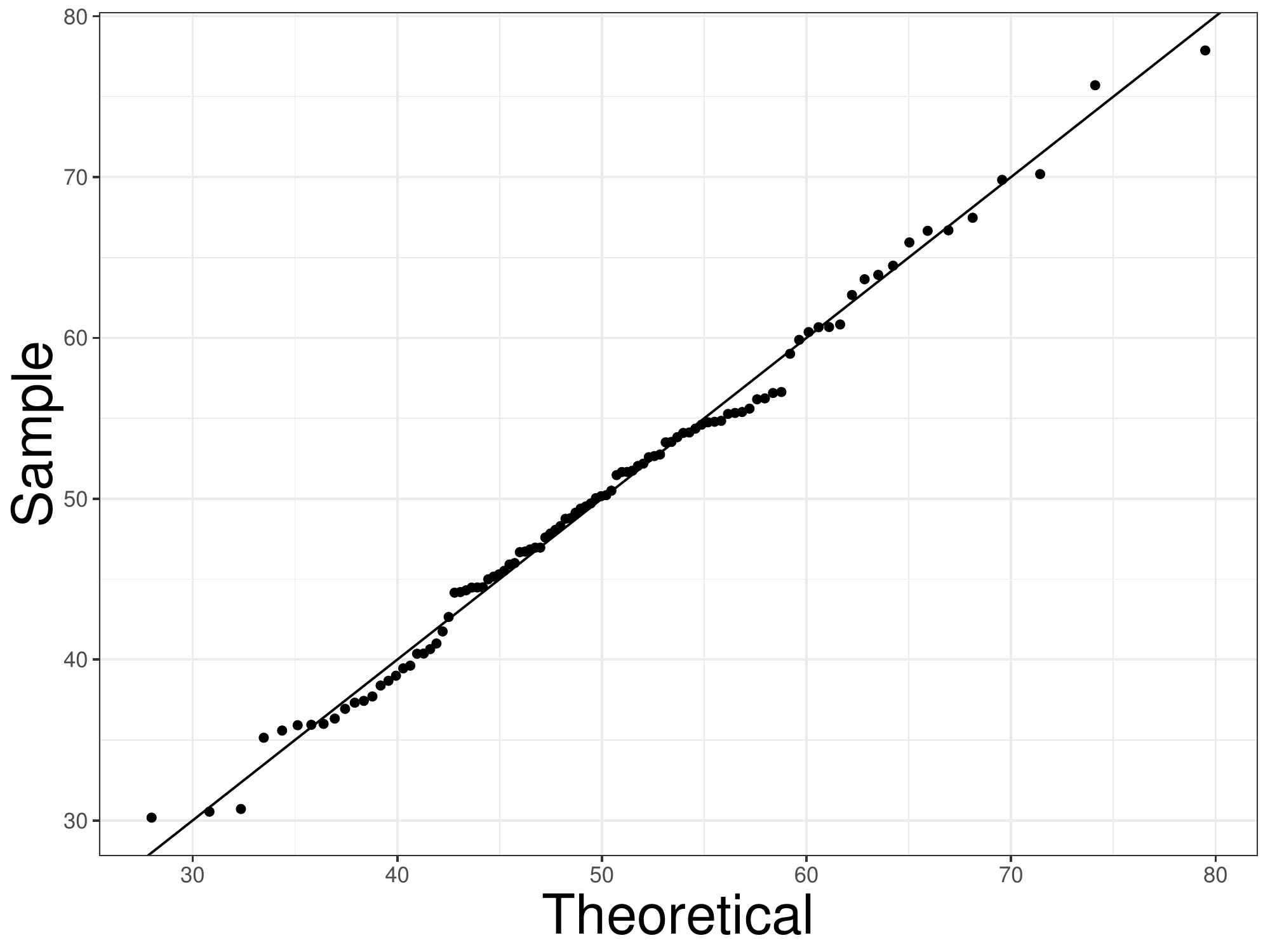}}
\subfigure[$\mbox{SNR}=0.1,\rho_x=0,\mbox{group}\ 2$]{\includegraphics[width=0.32\textwidth]{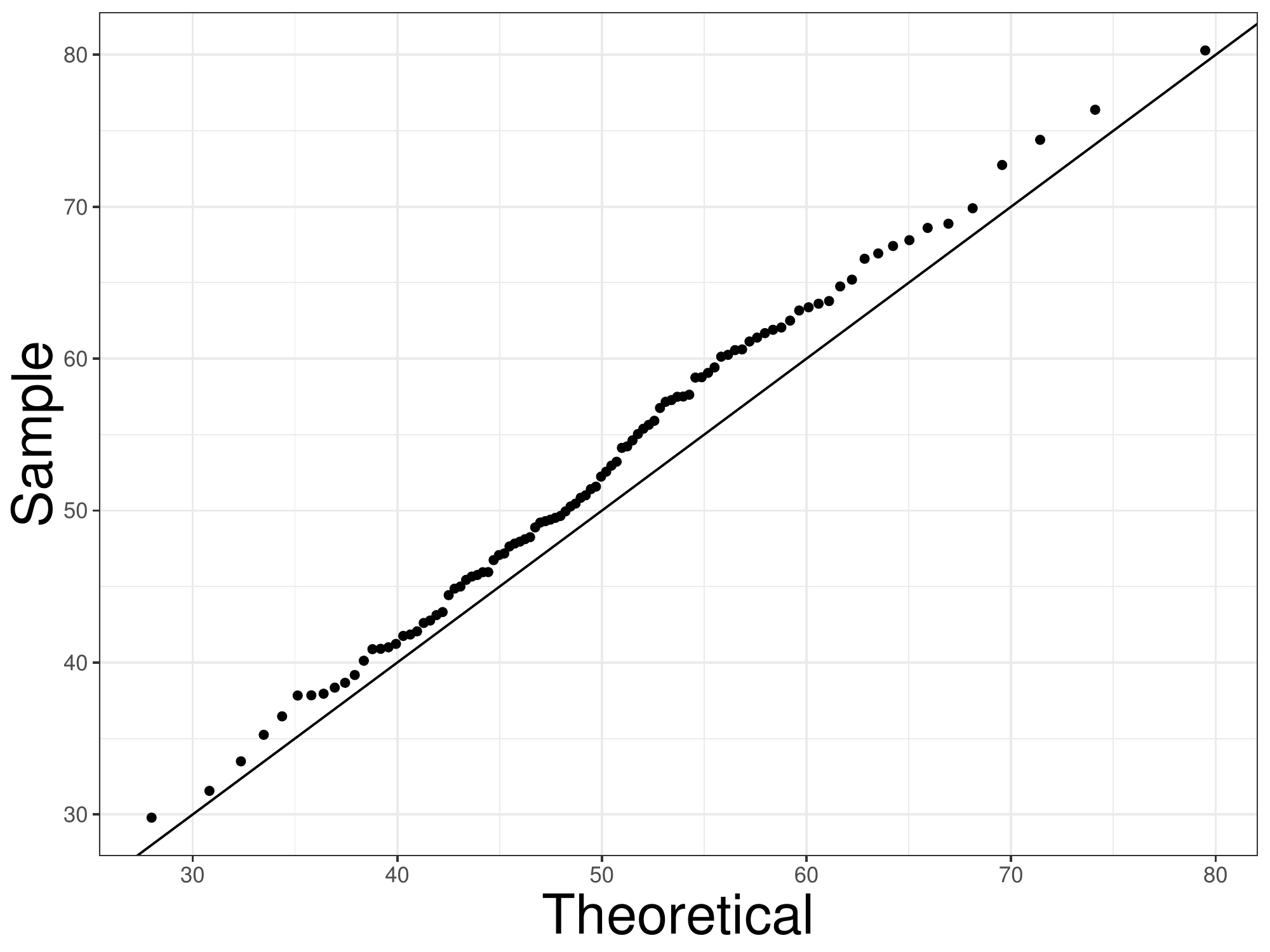}}
\subfigure[$\mbox{SNR}=0.1,\rho_x=0,\mbox{group}\ 3$]{\includegraphics[width=0.32\textwidth]{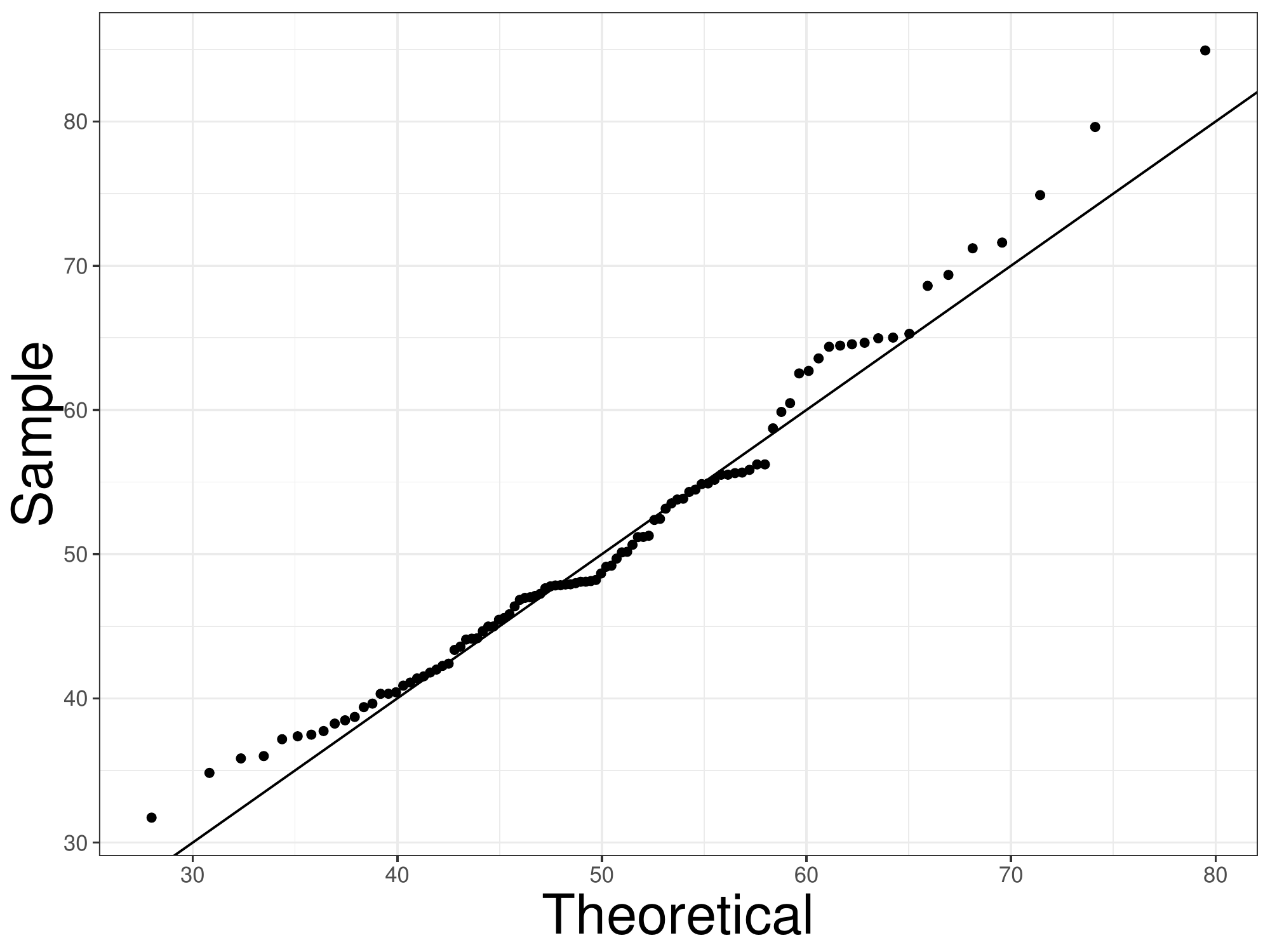}}
\caption{ Simulation results for setting 1 with $\rho_x=0$ from 100 simulation runs: (a) - (c) are the Q-Q plots of $\sqrt{2nq}(\hat\sigma/\sigma-1)$ versus $\mathcal{N}(0,1)$; (d) - (f) are the Q-Q plots of $\|\P_k\E - \mbox{Rem}_k\|_F^2/\hat\sigma^2$ versus $\chi^2_{r_k'q}$.}\label{fig:rho0}
\end{figure}

Table \ref{tab:s3} reports the detailed results on hypothesis testing under different settings with $\X$ being generated from the among-group correlation setup. Correlation patterns among covariates appear to have little effect on these results. See Table \ref{tab:within-group} for the results with the within-group correlation setup. In general, the magnitude of SNR and the model dimensionality have great influence on the TP, which measures the power of the test, while the type I error rate, i.e., FP, is not sensitive to the change of these two factors and only oscillates slightly around 0.05. Specifically, in Setting 1 where $n < p$, the power of the test is moderate when SNR is very low. When the SNR becomes stronger, the power of the test in these two settings dramatically increase to be close to 1. Setting 2 is a high-dimensional situation; when SNR is low, the power of the test is generally lower than in Setting 1. With a higher signal strength, i.e., SNR $=0.4$, the power of the test achieves 1. Moreover, a comparison between multivariate and univariate inference results demonstrate the power gaining by simultaneously modeling multiple responses, especially when SNR is low. %In summary, the simulation results indicate that when the true signal is very weak, the power of the test is not optimal, but a slight increase in signal strength can lead to a surge of the power. Also, under various situations, the probability of the occurrence of type I error can always be controlled around a pre-assigned level. 

\begin{table}[htbp]
\centering
\caption{Simulation results with $\X$ being generated from the among-group correlation setup. The performance of noise level estimation is displayed in terms of the mean ($\times 100$) and standard error ($\times 100$, in parenthesis) of $\hat\sigma/\sigma-1$ and $|\hat\sigma/\sigma-1|$, respectively. In both settings, we have $r_1^*=2$ and $r_2^*=r_3^*=0$. Each group is denoted as ``G'' followed by its group number.}\label{tab:s3}
\begin{tabular}{lccrclrcl}
  \hline
Design & \multirow{2}{*}{$\hat \sigma/\sigma-1$} & \multirow{2}{*}{$|\hat \sigma/\sigma-1|$} & \multicolumn{3}{c}{Multivariate Inference} & \multicolumn{3}{c}{Univariate Inference} \\
(SNR, $\rho_x$) &   & & G1      &     G2     &     G3 & G1      &     G2     &     G3\\
\hline
\multicolumn{9}{ c }{Setting 1}\\
\hline
(0.1,0.0) & -0.27 (1.68) & 1.36 (1.01) & 0.65 & 0.07 & 0.08 & 0.47 & 0.02 & 0.06 \\
(0.1,0.5) & -0.14 (1.68) & 1.36 (0.99) & 0.62 & 0.05 & 0.07 & 0.46 & 0.02 & 0.06 \\  
\hline
(0.2,0.0) & -0.72 (1.60) & 1.41 (1.04) & 1.00 & 0.07 & 0.09 & 1.00 & 0.03 & 0.07 \\
(0.2,0.5) & -0.64 (1.66) & 1.43 (1.06) & 1.00 & 0.04 & 0.07 & 1.00 & 0.03 & 0.07 \\ 
\hline
(0.4,0.0) & -0.94 (1.61) & 1.50 (1.09) & 1.00 & 0.08 & 0.09 & 1.00 & 0.04 & 0.09 \\
(0.4,0.5) & -0.96 (1.65) & 1.53 (1.13) & 1.00 & 0.06 & 0.08 & 1.00 & 0.03 & 0.07 \\ 
\hline 
\multicolumn{9}{ c }{Setting 2}\\
\hline
(0.1,0.0) & -0.15 (1.75) & 1.34 (1.13) & 0.16 & 0.05 & 0.03 & 0.11 & 0.02 &  0.02 \\ 
(0.1,0.5) & 0.04 (1.71) & 1.35 (1.05) & 0.16 & 0.05 & 0.03 & 0.17 & 0.03 &  0.03 \\ 
\hline
(0.2,0.0) & -0.50 (1.77) & 1.38 (1.20) & 0.79 & 0.05 & 0.03 & 0.80 & 0.02 & 0.02  \\
(0.2,0.5) & -0.10 (1.72) & 1.39 (1.01) & 0.74 & 0.07 & 0.05 & 0.76 & 0.03 & 0.04  \\
\hline
(0.4,0.0) & -0.66 (1.97) & 1.55 (1.38) & 1.00 & 0.06 & 0.04 & 1.00 & 0.03 & 0.03  \\  
(0.4,0.5) & -0.25 (2.07) & 1.64 (1.27) & 1.00 & 0.07 & 0.04 & 1.00 & 0.04 & 0.04  \\
\hline
  \end{tabular}
\end{table}

%onerow & -0.15 (1.75) & 1.34 (1.13) & 0.16 & 0.05 & 0.03 \\ 
%  onerow.1 & 0.04 (1.71) & 1.35 (1.05) & 0.16 & 0.05 & 0.03 \\ 
%  onerow.2 & -0.5 (1.77) & 1.38 (1.2) & 0.79 & 0.05 & 0.03 \\ 
%  onerow.3 & -0.1 (1.72) & 1.39 (1.01) & 0.74 & 0.07 & 0.05 \\ 
%  onerow.4 & -0.66 (1.97) & 1.55 (1.38) & 1 & 0.06 & 0.04 \\ 
%  onerow.5 & -0.25 (2.07) & 1.64 (1.27) & 1 & 0.07 & 0.04 \\ 

%X & 0.11 & 0.02 & 0.02 \\ 
%  X.1 & 0.17 & 0.03 & 0.03 \\ 
%  X.2 & 0.80 & 0.02 & 0.02 \\ 
%  X.3 & 0.76 & 0.03 & 0.04 \\ 
%  X.4 & 1.00 & 0.03 & 0.03 \\ 
%  X.5 & 1.00 & 0.04 & 0.04 \\
%

%# > rej_Bon_all[-1,]
%# [,1] [,2] [,3]
%# a0.1 r0
%# 0.12 0.03 0.03
%# a0.1 r0.5
%# 0.18 0.04 0.04
%# a0.2 r0.5
%# 0.77 0.04 0.05

\begin{table}[htbp]
\centering
\caption{Simulation results with $\X$ being generated from the within-group correlation setup. The performance of noise level estimation is displayed in terms of the mean ($\times 100$) and standard error ($\times 100$, in parenthesis) of $\hat\sigma/\sigma-1$ and $|\hat\sigma/\sigma-1|$, respectively. In both settings, we have $r_1^*=2$ and $r_2^*=r_3^*=0$. Each group is denoted as ``G'' followed by its group number.}\label{tab:within-group}
%\resizebox{\linewidth}{!}{%
\begin{tabular}{lccrclrcl}
  \hline
Design & \multirow{2}{*}{$\hat \sigma/\sigma-1$} & \multirow{2}{*}{$|\hat \sigma/\sigma-1|$} & \multicolumn{3}{c}{Multivariate Inference} & \multicolumn{3}{c}{Univariate Inference} \\
(SNR, $\rho_x$) &   & & G1      &     G2     &     G3 & G1      &     G2     &     G3\\
\hline
\multicolumn{9}{ c }{Setting 1}\\
\hline
(0.1,0.0) & -0.27 (1.68) & 1.36 (1.01) & 0.65 & 0.07 & 0.08 & 0.47 & 0.02 & 0.06 \\  
(0.1,0.5) & -0.10 (1.67) & 1.34 (0.99) & 0.63 & 0.06 & 0.07 & 0.48 & 0.03 & 0.08 \\ 
\hline
(0.2,0.0) & -0.72 (1.60) & 1.41 (1.04) & 1.00 & 0.07 & 0.09 & 1.00 & 0.03 & 0.07 \\  
(0.2,0.5) & -0.63 (1.64) & 1.41 (1.04) & 1.00 & 0.07 & 0.08 & 1.00 & 0.04 & 0.08 \\ 
\hline
(0.4,0.0) & -0.94 (1.61) & 1.50 (1.09) & 1.00 & 0.08 & 0.09 & 1.00 & 0.04 & 0.09\\  
(0.4,0.5) & -0.95 (1.65) & 1.54 (1.11) & 1.00 & 0.08 & 0.08 & 1.00 & 0.04 & 0.07 \\ 
\hline
\multicolumn{9}{ c }{Setting 2}\\
\hline
(0.1,0.0) & -0.15 (1.75) & 1.34 (1.13) & 0.16 & 0.05 & 0.03 & 0.11 & 0.02 &  0.02 \\ 
(0.1,0.5) & 0.06 (1.68) & 1.32 (1.03) & 0.16 & 0.04 & 0.03 & 0.16 & 0.04 & 0.03 \\ 
\hline
(0.2,0.0) & -0.50 (1.77) & 1.38 (1.20) & 0.79 & 0.05 & 0.03 & 0.80 & 0.02 & 0.02 \\ 
(0.2,0.5) & -0.12 (1.73) & 1.40 (1.01) & 0.78 & 0.05 & 0.03 & 0.76 & 0.03 & 0.03 \\ 
\hline
(0.4,0.0) & -0.66 (1.97) & 1.55 (1.38) & 1.00 & 0.06 & 0.04 & 1.00 & 0.03 & 0.03\\ 
(0.4,0.5) & -0.23 (1.99) & 1.59 (1.21) & 1.00 & 0.04 & 0.05 & 1.00 & 0.03 & 0.04 \\  
\hline
 \end{tabular}%}
\end{table}

%# 0.48 0.03 0.08
%# 1.00 0.04 0.08
%# 1.00 0.04 0.07

%# 0.16 0.04 0.03
%# 0.76 0.03 0.03
%# 1.00 0.03 0.04

% Under the significance level 0.05, the designed test has false positive rates around 0.07 under setting 1. 
 %For testing the first view, which is indeed predictive to the response, the true positive rate is 1 for small and moderate signal strengths, i.e., SNR $=0.2$ or 0.4. Only for the extremely weak signal with SNR $=0.1$, the true positive rate is less than 1 and is around 0.6. 
 %The test results are similar under setting 2. When the signal strength is extremely weak, the true positive rate is low and when SNR $=0.2$ or 0.4, the power of the rest can achieve 1. 
 %For setting 3, the high-dimensional case where the sample size is much less than the number of predictors, when SNR $=0.1$ the true positive rate is around 0.1, when SNR $=0.2$ the true positive rate is around 0.7 and with a moderate signal strength SNR $=0.4$ the true positive rate is 1. 
 %This indicates that when the true signal is itself very weak and severely messed up by random noises, the proposed procedure cannot precisely identify the predictive views but even under this situation, the false positive rate will not deviate from 0.05 too much. So the inference procedure can still give us reliable conclusions. 

\subsection{Simulation with Generated Compositional Data}

%Considering the motivation of the proposed multivariate log-contrast model and its inference procedure, 
We conduct simulations based on generated compositional data with a similar setting as the preterm infant dataset. Specifically, we let $n=40$, $p=60$, $q=10$, $K=10$ and each group is of size 6. Similar to \citet{shi2016regression}, we obtain vectors of count $\w_i=(\w_{1,i}\trans,\ldots,\w_{K,i}\trans)\trans\in\mathbb{R}^p,\ i=1,\ldots,n$ from a log-normal distribution $\ln\mathcal{N}_p(\bmu^w,\bSigma^w)$. In order to reflect the difference in abundance of each taxon in the microbiome counts observation, we let %the first five groups each contain a taxon with a higher abundance, i.e.,  
$\bmu_{k}^w=(10,1,1,1,1,1)\ \text{for}\ k=1,\ldots,5 $ and $\bmu_{k}^w=(1,1,1,1,1,1)\ \text{for}\ k=6,\ldots,10,$ where $\bmu_{k}^w$ is the mean vector corresponding to the $k$-th group. To simulate the commonly existing correlation among counts of taxa, we use $\bSigma^w=(\rho_x^{|i-j|})$ with $\rho_x \in \{0.2, 0.5\}$. We then transform the count data into sub-compositional data, i.e.,
\begin{align*}
	z_{k,i,j}=\frac{w_{k,i,j}}{\sum_{j=1}^{p_k}w_{k,i,j}},\ k=1\ldots,K;\ i=1,\ldots,n,
\end{align*}
where $w_{k,i,j}$ is the count of the $j$-th taxon within the $k$-th group of the $i$-th subject. 
 
To see the potential effect of the existence of highly abundant taxon on the group inference results, %Since two kinds of within group density distributions are considered, from each of them we select one predictive group to examine the possible difference in estimating and testing performance caused by it. Thus, we 
we select the first and the sixth group as two predictive groups, and let $\mbox{rank}(\B_1^*)=\mbox{rank}(\B_6^*)=2$ with all the other groups have zero sub-coefficient matrices. %Each $\C_k^*$ is generated as before and satisfies $\1_{p_k}\trans\C_k^*=\0$. %in the same way as before with one difference that the first $p_k-1$ rows of $\J_k$ are drawn from a standard normal distribution and the last row of it comes from the restriction $\1_{p_k}\trans\J_k=\0$. 
Without further scaling to $\B_k^*$ we obtain $\Y$ from (\ref{sub:nores}) %where the random error is generated in the same way as before with the noise level is adjusted by altering 
with SNR $\in\{1,2,4\}$.

The results are shown in Table \ref{comp_simu1}, where we report the noise level estimation results and the testing results for two predictive groups and two irrelevant groups based on 100 replications. 
%Different from the underestimation of the noise level displayed in Table \ref{tab:s1} -- \ref{tab:s3}, based on the generated compositional data, the scaled iRRR leads to overestimation of $\sigma$, which becomes most obvious when SNR $=4$. 
In general, the power of the test is relatively low when the signal is weak (SNR $=1$), and it increases to 1 swiftly when the SNR becomes larger. The false positive rate is controlled around 0.05 for all situations. In particular, whether or not a group contains highly abundant taxa has a large impact on the test power. % the proposed test has poor performance on the groups that contain taxon with high abundance. %As for the two irrelevant groups, the false positive rate among tests for group 2 is slightly less than the one for group 8. 
The power of the test for the first group is much less than the power of the test for the sixth group when the signal is weak. This phenomenon could be related to $d_1(\P_k(\I_n-\Q_k))$ that is directly affected by the abundance level. Specifically, in this simulation, for the groups that have unbalanced taxa distributions, their $d_1(\P_k(\I_n-\Q_k))$ is close to 1, and for the groups whose components take comparable proportions, the corresponding $d_1(\P_k(\I_n-\Q_k))$ is around 0.8. As we discussed before, $d_1(\P_k(\I_n-\Q_k))$ measures the uniqueness of the information carried by the $k$-th group, thus a smaller value indicates a higher inference accuracy. In addition, we can see a great improvement in the power brought by the multivariate analysis when $d_1(\P_k(\I_n-\Q_k))$ is large.

\begin{table}[ht]
\centering
\caption{Simulation results based on the generated compositional data across 100 replications. The performance of noise level estimation is displayed in terms of the mean ($\times 10$) and standard error ($\times 10$, in parenthesis) of $\hat\sigma/\sigma-1$ and $|\hat\sigma/\sigma-1|$, respectively. In the simulation setting, we have $r_1^*=r_6^*=2$ and $r_2^*=r_7^*=0$. Each group is denoted as ``G'' followed by its group number.}\label{comp_simu1}
\resizebox{\linewidth}{!}{%
\begin{tabular}{lccrrrrrrrr}
\hline
Design & \multirow{2}{*}{$\hat \sigma/\sigma-1$} & \multirow{2}{*}{$|\hat \sigma/\sigma-1|$} & \multicolumn{4}{c}{Multivariate Inference} & \multicolumn{4}{c}{Univariate Inference} \\
(SNR, $\rho_X$) &   &   &     G1      &     G2  & G6    &     G7 &     G1      &     G2  & G6    &     G7\\
\hline
(1, 0.2) & 0.44 (0.47) & 0.51 (0.39) & 0.36 & 0.03 & 0.67 & 0.03 & 0.10 & 0.01 & 0.63 & 0.04  \\ 
(1, 0.5) & 0.29 (0.47) & 0.42 (0.35) & 0.25 & 0.02 & 0.62 & 0.02 & 0.06 & 0.01 & 0.67 & 0.04 \\ 
(2, 0.2) & 1.60 (0.73) & 1.60 (0.72) & 0.93 & 0.02 & 1.00 & 0.01 & 0.69 & 0.01 & 1.00 & 0.04 \\ 
(2, 0.5) & 1.01 (0.60) & 1.01 (0.60) & 0.88 & 0.02 & 1.00 & 0.01 & 0.50 & 0.02 & 1.00 & 0.03 \\ 
(4, 0.2) & 5.70 (1.50) & 5.70 (1.50) & 1.00 & 0.00 & 1.00 & 0.00 & 0.97 & 0.01 & 1.00 & 0.02 \\ 
(4, 0.5) & 3.83 (1.11) & 3.83 (1.11) & 1.00 & 0.00 & 1.00 & 0.01 & 0.95 & 0.00 & 1.00 & 0.03 \\ 
\hline
\end{tabular}}
\end{table}

\section{Assessing the Association Between Preterm Infants' Gut Microbiome and Neurobehavioral Outcomes}\label{sec:app}
%{\color{red} I hope we can do some more analysis. (1) How about we split the entire period (about 30 days) to three periods (10 days each), and do analysis for each sub-period. The overall analysis will still be kept. In this way, we can look at how the impact of microbiomes changes over the period. We can argue that the data is sparse, so taking averages for each 10 day period is a good compromise. (2) We may do some univariate analysis, to show the power we gained by doing multivariate analysis.}
\subsection{Data Description}

The study was conducted at a Level IV NICU in the northeast region of the U.S. Fecal samples were collected in a daily manner when available during the first month of the postnatal age of infants, from which bacteria DNA were isolated and extracted \citep{bomar2011directed, cong2017influence,sun2018log}. The V4 region of the 16S rRNA genes was sequenced and analyzed using the Illumina platform and QIIME \citep{cong2017influence}, and microbiome data were obtained. There were $n=38$ infants under study.

%: the lower the rank, the higher the resolution of the taxonomic categories, but the sparser the data for each category. 
%A good compromise is achieved by the sub-compositional regression analysis \citep{shi2016regression}, in which the effect of a taxon on the outcome at the rank of primary interest is investigated through its more information-rich sub-compositions at a lower taxonomic rank.

%In order to conduct the analysis on the same footing as \citet{sun2018log}, who first analyzed this dataset through a sparse log-contrast functional regression method to identify predictive bacteria categories to a single neurodevelopment measurement,

In practice, the selection of the taxonomic ranks at which to perform the statistical analysis depends on both the scientific problem itself and trade-off between data quality and resolution: the lower the taxonomic rank, the higher the resolution of the taxonomic units, but the sparser or the less reliable the data in each unit. To achieve a compromise, here we perform a sub-compositional analysis: we assess the effects of the order-level gut microbes through compositions at the genus level, a lower taxonomic rank. The microbes were categorized into 62 genera ($\sum_{k=1}^K p_k=62$), which can be grouped into $K=11$ predictor sets based on their orders. The original orders only containing a single genus were put together as the ``Other'' group.  %Here we remark that, although both \citet{sun2018log} and our model aim to identify predictive orders, \citet{sun2018log} used the summation of all genera counts under one certain order as a predictor and our method takes into consideration the collaboration effect of genera and dissects the problem from a finer scale. Moreover, instead of modeling the longitudinal gut microbiota observations as functionals as in \citet{sun2018log}, here 
The preterm infant data were longitudinal, with on average $11.4$ daily observations per infant through the 30 day postnatal period. In this study, we concern average microbiome compositions in three stages, i.e., stage 1 (postnatal age of 0-10 days, $n=33$), stage 2 (postnatal age of 11-20 days, $n=38$) and stage 3 (postnatal age of 21-30 days, $n=29$), in order to enhance data stability and capture the potential time-varying effects of the gut microbiome on the later neurodevelopmental responses. We also performed analysis on the average compositions of the entire time period. Figure \ref{fig:baby} displays the average abundances of the orders for each infant at different stages. Before calculating the compositions, we replaced the zero counts by 0.5, the maximum rounding error, to avoid singularity \citep{aitchison2003statistical}. Several control variables characterizing demographical and clinical information of infants were included ($p_0=6$), including gender (binary, female $=$ 1), delivery type (binary, vaginal $=$ 1), premature rupture of membranes (PROM, yes $=$ 1), score for Neonatal Acute Physiology-Perinatal Extension-II (SNAPPE-II), birth weight (in gram) and the mean percentage of feeding by mother's breast milk (MBM, in percentage). 

\begin{figure}[htp]
\centering 
\includegraphics[width=0.9\columnwidth]{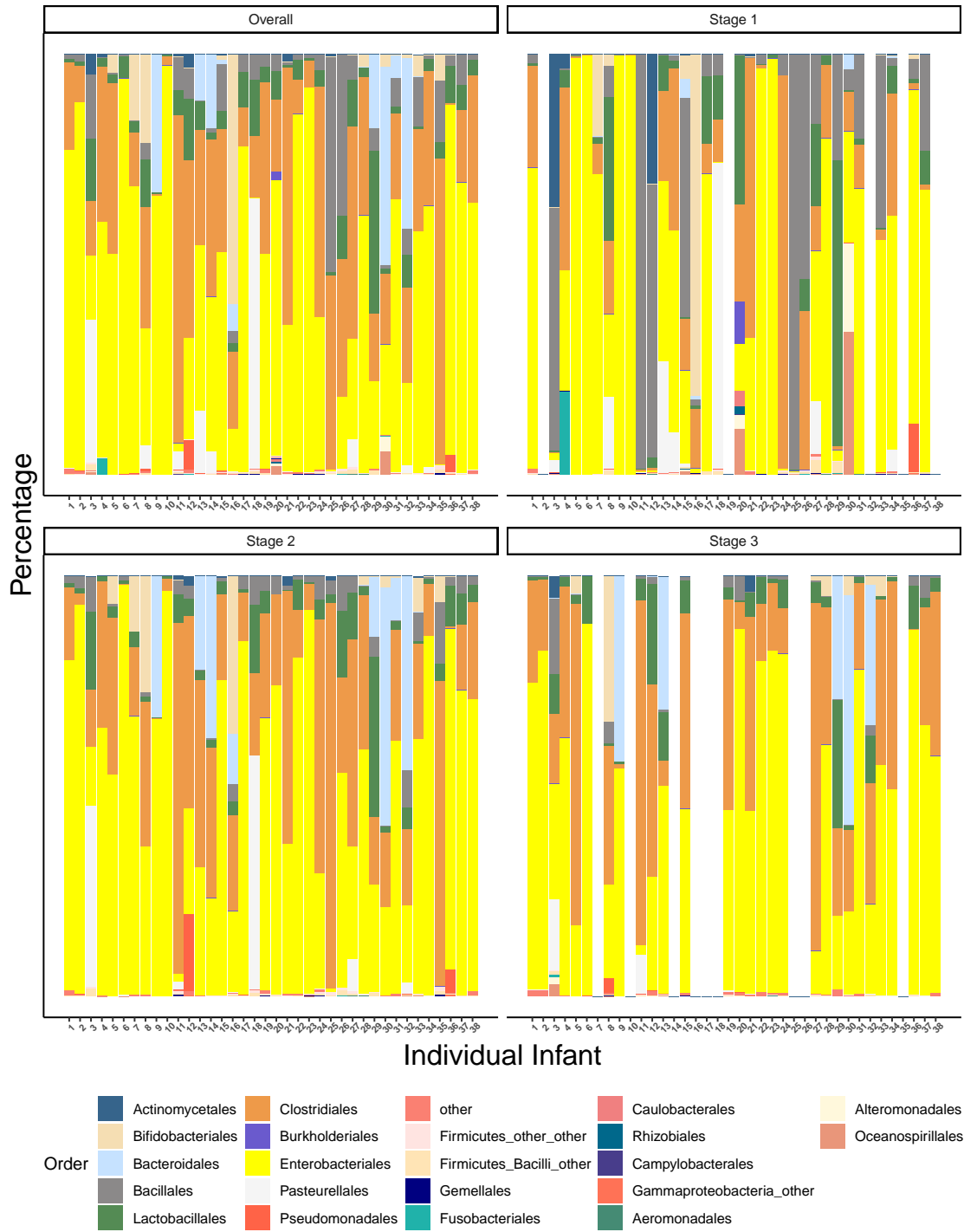}
\caption{The average abundance profiles of the 22 orders for each infant at the three stages: stage 1 (postnatal age of 0--10 days; $n=33$), stage 2 (postnatal age of 11--20 days; $n=38$) and stage 3 (postnatal age of 21--30 days; $n=29$). The profiles of the infants with no observation are shown as in white color.  
}\label{fig:baby}
\end{figure}

The infants' neurobehavioral outcomes were measured when the infant reached 36-38 weeks of gestational age using NNNS. NNNS is a comprehensive assessment of both neurologic integrity and behavioral function for infants. It consists of 13 sub-scale scores including habituation, attention, handling, quality of movement, regulation, nonoptimal reflexes, asymmetric reflexes, stress/abstinence, arousal, hypertonicity, hypotonicity, excitability and lethargy. These scores were obtained by summarizing several examination results within each sub-category in the form of the sum or mean, and all of them can be regarded as continuous measurements with a higher score on each scale implying a higher level of the construct \citep{lester2004neonatal}. %For example, a higher score on stress/abstinence means that the infant was more stressful.
We discarded sub-scales hypertonicity and asymmetric reflexes since their scores are $90\%$ zero and focused on the other 11 standardized sub-scale scores ($q=11$).

\subsection{Results}   

%\begin{table}[ht]
%\centering
%\caption{Sub-Compositional Analysis with $q=11$ (P-values Corrected with BH Adjustment)}\label{tab:BHapp2}
%\begin{tabular}{lrrrrr}
%  \hline
% Order & P-value(s0) & P-value(s1) & P-value(s2) & P-value(s3) & FuncResults \\ 
%  \hline
%  \hline
%Actinomycetales & 0.850 & 0.498 & 0.708 & {\color{red}0.019} & F \\ 
%  Bifidobacteriales & 0.888 & 0.498 & 0.934 & 0.907 & F \\ 
%  Bacteroidales & 0.850 & 0.498 & 0.762 & 0.752 & F \\ 
%  Bacillales & 0.850 & 0.498 & 0.573 & 0.707 & F \\ 
%  Lactobacillales & 0.102 & 0.498 & 0.193 & 0.785 & {\color{red}T} \\ 
%  Clostridiales & {\color{red}0.018} & 0.401 & {\color{red}0.059} & 0.337 & {\color{red}T} \\ 
 % Burkholderiales & 0.850 & 0.498 & {\color{red}0.000} & 0.707 & F \\ 
 % Enterobacteriales & 0.397 & 0.498 & 0.845 & 0.752 & {\color{red}T} \\ 
 % Pasteurellales & 0.397 & 0.498 & 0.934 & 0.752 & F \\ 
 % Pseudomonadales & 0.850 & 0.498 & 0.821 & 0.752 & F \\ 
 % Others & {\color{red}0.004} & 0.498 & 0.150 & 0.500 & {\color{red}T} \\ 
 %  \hline
%\end{tabular}
%\end{table}

The results are shown in Table \ref{tab:BHapp2}. To control the false discovery rate (FDR) when multiple tests are conducted, we mark the orders based on the corrected $p$-values with Benjamini-Hochberg adjustment \citep{benjamini1995controlling}. %where we report the results obtained from three time-specific models and also the overall model (the uncorrected p-values are listed in Supplemental Material). 
%We use the abbreviations s0, s1, s2 and s3 to denote the overall model, stage 1 model, stage 2 model and stage 3 model, respectively. 
First, we observe that the predictive effects of the microbiome on the neurobehavioral development measurements appear to be dynamic, i.e., in different time periods, the identified taxa are not the same. This reflects the fact that the gut microbiome compositions in early postnatal period are highly variable, due to their sensitivity to illnesses, changes in diet and environment \citep{nuriel2016microbial, koenig2011succession}. %Illnesses, changes in diet and environment could greatly affect the gut microbiome composition before it becomes mature \citep{koenig2011succession}. 
{Specifically, by controlling the FDR under 10\%, the identified orders from all analyses are Actinomycetales, Clostridiales, Burkholderiales and the aggregated group ``Others''. If we set 0.05 as the significance level without multiple testing adjustment, there is one more significant order, Lactobacillales. %Clostridiales are detected at stage 1; Clostridiales, Burkholderiales and the aggregated group `Others' are identified at stage 2; and Actinomycetales are selected at stage 3. From the global level analysis, Lactobacillales, Clostridiales and the aggregated group `Others' are selected. 
  This dataset is also analyzed by \citet{sun2018log} through a sparse log-contrast functional regression method to identify predictive gut bacterial orders to the stress/abstinence sub-scale, and their selected orders based on penalized estimation include Lactobacillales, Clostridiales, Enterobacteriales and the group ``Others'', which are very consistent with our results. Here we stress that our work is quite different from \citet{sun2018log}: our analysis assesses the multivariate association between the orders and the multiple neurodevelopment measurements using a valid statistical inference procedure through sub-compositional analysis, while \citet{sun2018log} emphasized estimating the dynamic effects of the orders to the stress score alone by fitting a regularized functional regression with the order-level data.}

\begin{table}[ht]
\centering
\caption{Raw p-values from the sub-compositional analysis applied to the preterm infant data. Without multiple adjustment, the identified orders under significance level 0.05 are marked in bold. With BH adjustment to control the FDR under 10\%, the identified orders are marked with an asterisk.}\label{tab:BHapp2}
\begin{tabular}{lllll}
  \hline
 Order & Overall & Stage 1 & Stage 2 & Stage 3 \\ 
  \hline
  \hline
Actinomycetales & 0.55 & 0.50 & 0.38 & \textbf{0.00}$^*$ \\ 
  Bifidobacteriales & 0.89 & 0.37 & 0.85 & 0.91 \\ 
  Bacteroidales & 0.70 & 0.49 & 0.49 & 0.62 \\ 
  Bacillales & 0.77 & 0.34 & 0.26 & 0.32 \\ 
  Lactobacillales & \textbf{0.03} & 0.18 & 0.07 & 0.71\\ 
  Clostridiales & \textbf{0.00}$^*$ & \textbf{0.04} & \textbf{0.01}$^*$ & 0.06 \\ 
  Burkholderiales & 0.47 & 0.45 & \textbf{0.00}$^*$ & 0.26 \\ 
  Enterobacteriales & 0.18 & 0.12 & 0.69 & 0.57 \\ 
  Pasteurellales & 0.15 & 0.26 & 0.93 & 0.50  \\ 
  Pseudomonadales & 0.64 & 0.25 & 0.60 & 0.61 \\ 
  Others & \textbf{0.00}$^*$ & 0.43 & \textbf{0.04} & 0.14 \\ 
   \hline
\end{tabular}
\end{table}

We have also conducted univariate analysis for each sub-scale, i.e., we fit the proposed model with each sub-scale score as the univariate response and make inference. For each time period, this procedure produces a large number of tests. By controlling the FDR under 10\%, only in stage 2 we can identify the order Burkholderiales to be predictive to excitability, stress/abstinence and handling. %(see Figure \ref{fig:BHP_facet}). 
This is not surprising as this dataset has a limited sample size and a weak signal strength. To gain more power in practice, the proposed multivariate test can be used first to verify the existence of any association between neurodevelopment and gut taxa, and univariate tests can then be conducted as post-hoc analysis to further inspect the pairwise associations. As such, we only conduct the univariate tests related to the orders identified from our multivariate analysis. With the FDR controlled at 10\%, two significant orders can be found in stage 2, i.e., Clostridiales and Burkholderiales (see Table \ref{app:post}).

% \begin{figure}[htp]
% \centering 
% \includegraphics[width=1\columnwidth]{BHP_facet.pdf}
% \caption{The identified predictive orders for each sub-scale score of NNNS when control the FDR under 10\% for each time-specific analysis. The selected orders are marked in red, while the remaining orders are marked in pink. }\label{fig:BHP_facet}
% \end{figure}

\begin{table}[ht]
\centering
\caption{Corrected p-values from the univariate analysis adjusted by using BH adjustment. For each stage, we control the FDR of the tests related to the orders identified in multivariate analysis based on the corrected p-values. The values highlighted with an asterisk are the significant ones by controlling the FDR under 10\%.}\label{app:post}
\resizebox{\linewidth}{!}{%
\begin{tabular}{llllll}
\hline
& \multicolumn{2}{c}{Overall} & \multicolumn{2}{c}{Stage 2} & Stage 3 \\
\hline
& Clostridiales & Others & Clostridiales & Burkholderiales & Actinomycetales \\ 
\hline
Habituation & 0.302 & 0.265 & 0.733 & 0.263 & 0.122 \\ 
Attention & 0.260 & 0.260 & 0.080$^*$ & 0.316 & 0.291 \\ 
Handling & 0.536 & 0.611 & 0.733 & 0.006$^*$ & 0.688 \\ 
Qmovement & 0.370 & 0.260 & 0.388 & 0.117 & 0.510 \\ 
Regulation & 0.260 & 0.260 & 0.212 & 0.054$^*$ & 0.111 \\ 
Nonoptref & 0.588 & 0.260 & 0.651 & 0.733 & 0.854 \\ 
Stress & 0.260 & 0.506 & 0.316 & 0.006$^*$ & 0.111 \\ 
Arousal & 0.260 & 0.260 & 0.212 & 0.214 & 0.831 \\ 
Hypotone & 0.260 & 0.260 & 0.252 & 0.893 & 0.131 \\ 
Excitability & 0.260 & 0.260 & 0.316 & 0.003$^*$ & 0.111 \\ 
Lethargy & 0.519 & 0.260 & 0.212 & 0.733 & 0.831 \\ 
\hline
\end{tabular}}
\end{table}

%While multivariate analysis tries to verify the existence of any association between neurodevelopment and gut taxa, the univariate analysis serves as a post-hoc test tool to inspect the pairwise associations. % even less 
%This shows the boosted power by conducting multivariate analysis. To see which sub-scales are affected by the identified orders from the multivariate analysis,  
%}

%  With a reference to the microbial analysis literature, some

Most of the identified orders are known to be of various biological functions to human beings. Both Lactobacillales and Clostridiales belong to the phylum Firmicutes, which are found to be abundant for infants fed with mother's breast milk \citep{cong2016gut}. Lactobacillales are usually found in decomposing plants and milk products, %and produce lactic acid as the major metabolic end-product of carbohydrate fermentation. Moreover, 
and they commonly exist in food and are found to contribute to the healthy microbiota of animal and human mucosal surfaces. 
Clostridiales are commonly found in the gut microbiome and some Clostridiales-associated bacterial genera in the gut are correlated with brain connectivity and health function \citep{labus2017clostridia}. The order Burkholderiales includes pathogens that are threatening to intensive health care unit patients and lung disease patients \citep{voronina2015variability}. It is also related to inflammatory bowel disease, especially for children's ulcerative colitis \citep{rudi2012analysing}.
%, and some of its species are known to be threatening to patients with underlying lung diseases \citep{hauser2011clinical}.
%On the stage 3 model, the generally gram-positive and anaerobic order Actinomycetales is identified to be significantly effective. As a heterogeneous order, 
{%Actinomycetales contain a wide range of bacteria, with some of them are located to be the cause of several infectious diseases. 
The genus Actinomyces from the order Actinomycetales is observed in this study. %, including Actinomyces, Varibaculum and Corynebacterium. %, where a variety of research has been conducted on Actinomyces. 
As a commensal bacteria that colonizes the oral cavity, gastrointestinal or genitourinal tract,  Actinomyces normally cause no disease. However, invasive disease may occur when mucosal wall undergoes destruction \citep{gillespie2014medical}. Moreover, certain species in Actinomyces is known to possess the metabolic potential to breakdown and recycle organic compounds, e.g., glucose and starch \citep{hanning2015functionality}.} %{\color{red} Make sure your discussion is related to human!}%  
As for the effects of the control variables on the neurodevelopment of preterm infants, the estimated coefficients from the overall model and its related discussion are in Appendix \ref{web:G}.

To summarize, the identified bacterial taxa are mostly consistent with existing studies and biological understandings. Therefore, our approach provides rigorous supporting evidence that stressful early life experiences imprint gut microbiome through the regulation of the gut-brain axis and impact later neurodevelopment.

\section{Discussion}\label{sec:fut}

We propose a multivariate multi-view log-contrast model to facilitate sub-composition selection, which together with an asymptotic hypothesis testing procedure successfully boosts the power in identifying neurodevelopment related bacteria taxa in the preterm infant study.  
%We propose a multivariate log-contrast regression model to assess the association between multiple clinical outcomes and the bacterial taxa considering the taxonomic hierarchy of bacteria. 
%The corresponding estimation and inference method are also provided to enable group-wise variable selection. 
%Specifically, we generalize the low-dimensional projection estimator to correct the bias of the estimators obtained from the scaled iRRR method to facilitate statistical inference.
%In the application to a preterm infant gut microbiome study, we successfully identifies some predictive microbes to neurodevelopment outcomes, which provides a set of potential biomarkers to some neurological diseases of preterm infants. 
{There are many directions for future research. We use cross validation to select the tuning parameter in the scaled iRRR, and in most situations the over-selection of cross validation leads to underestimation of the noise level and inflation of the false positive rate in the subsequent inference. We will explore other approaches of tuning to obtain a more accurate estimate of the noise level. Another pressing issue is to investigate the robustness of the method to the violation of the homoscedasticity, independence, and normality of the error terms, since the theoretical guarantees of the scaled iRRR and the inference procedure are built on these strong assumptions. The extension of the method to deal with multivariate Non-Gaussian response is also an interesting topic due to the widely existing binary or count responses in practical applications.} Moreover, it is worthwhile to generalize the proposed method to longitudinal microbiome data analysis. In the preterm infant study, it is interesting to model the dynamic effect of a bacteria taxon on neurodevelopment measurements as a function of time and make inference on it. 

%\if1\blind
%{
%\section*{Acknowledgement}
%
%}\fi

\appendix
\clearpage
%\begin{centering}
  {\centering\bf\huge Appendix}
%\end{centering}

\section{Computation Details}\label{web:A}

In this section, computational algorithms are introduced to obtain the scaled iRRR estimator and the score matrix. For readers' convenience, we first reproduce the scaled composite nuclear norm penalization approach \eqref{eq:main2} and the score matrix estimation framework \eqref{inf8} of the main paper. They are given by  
\begin{align}
(\widehat \bmu, \widehat \C_0, \widehat \B^n, \hat \sigma ) & = \arg \min_{\bmu, \C_0, \B, \sigma} \L_{\w}(\bmu, \C_0, \B, \sigma) \nonumber \\
& = \arg \min_{\bmu, \C_0, \B, \sigma} \left\{ \frac{1}{2nq \sigma} \|\Y - \1_n\bmu\trans - \Z_0\C_0 - \X\B\|_F^2 + \frac{\sigma}{2} +\lambda\sum_{k=1}^K w_k \|\B_k\|_* \right\}\label{eq:main2supp0}	
\end{align}
and
\begin{align}
	\widehat \bGamma_{-k} = \arg\min_{\bGamma_{j, j\neq k}} \left\{ 
	\frac{1}{2n} \|\X_k - \sum_{j \neq k} \X_j \bGamma_j \|_F^2 + \sum_{j \neq k} \frac{\xi w_j^{''}}{\sqrt{n}} \|\X_j \bGamma_j\|_* \right\}, \label{inf8supp}
\end{align}
respectively. Since the intercept and the control variables can be treated as a group with penalty zero, instead of solving \eqref{eq:main2supp0}, we only need to focus on  
\begin{align}
(\widehat \B^n, \hat \sigma ) & = \arg \min_{\B, \sigma} \L_{\w}(\B, \sigma) \nonumber \\
& = \arg \min_{\B, \sigma} \left\{ \frac{1}{2nq \sigma} \|\Y -\X\B\|_F^2 + \frac{\sigma}{2} +\lambda\sum_{k=1}^K w_k \|\B_k\|_* \right\}.\label{eq:main2supp}	
\end{align}
 As for the two algorithms to solve (\ref{eq:main2supp}), one is derived as a block-wise coordinate descent algorithm and another is built on the alternating direction method of multipliers \citep[ADMM]{boyd2011distributed}. Both methods have good performance in our simulation. %, with the block-wise coordinate descent method being more computationally efficient. 
 As for the score matrix estimation with (\ref{inf8supp}), an ADMM algorithm is proposed.

\subsection{Scaled iRRR Estimation}\label{web:A1}

%{\color{red} Which method is used in your numerical study? It is strange to present two methods without discussing their pros and cons. Please present the one that is more justifiable (perhaps they are equally so) which should also be the one that is actually used.}

With a given $\sigma$, we have
\begin{align}
\sigma \L_{\w}(\B,\sigma)=\L_{\w^*}(\B)+\frac{\sigma^2}{2} \nonumber	
\end{align}
where $\w^*=\sigma \w = (\sigma w_1, \ldots, \sigma w_K)\trans$ and $\L_{\w^*}(\B)$ is the objective function in the original iRRR estimation framework \citep{li2018integrative}
\begin{align}
\widehat \B^n(\w) = \arg \min_{\B \in \mathbb{R}^{p \times q}} \L_{\w}(\B)= \arg \min_{\B \in \mathbb{R}^{p \times q}} \left\{ \frac{1}{2nq} \|\Y - \X\B\|_F^2+\lambda\sum_{k=1}^K w_k \|\B_k\|_* \right\}. \label{eq:main1}	
\end{align}
The notation $\widehat \B^n(\w)$ emphasizes the dependence of the estimator on the weight $\w$.
Therefore, a block-wise coordinate descent algorithm can be applied to solve (\ref{eq:main2supp}). Suppose at the $k$-th iteration, we have $\hat \sigma^{(k)}$ and $\widehat \B^{n(k)}(\w^{(k)})$. %For simplicity, here we write $\widehat \B^{n(k)}(\w^{(k)})$ as $\widehat \B^{n(k)}$. 
Then at the $(k+1)$-th iteration, %we first use the latest coefficient matrix estimate from the $k$-th iteration to update $\sigma$, then based on $\sigma^{(k+1)}$ to get an adjusted set of weights, and finally update the coefficient matrix by using the new weights. 
the updating procedure is summarized as
\begin{align}
&\hat \sigma^{(k+1)} \leftarrow \|\Y-\X\widehat\B^{n(k)}(\w^{(k)})\|_F/\sqrt{nq}, \label{eq:ite0} \\
&\w^{(k+1)} \leftarrow \w \hat \sigma^{(k+1)}, \\
&\widehat \B^{n(k+1)}(\w^{(k+1)}) \leftarrow \arg\min_{\B \in \mathbb{R}^{p\times q}} \L_{\w^{(k+1)}}(\B).  \label{eq:ite1}
\end{align}
%The updating formula (\ref{eq:ite0}) is obtained by using the fact that the profile objective function $\L_w(\widehat \B(\sigma w), \sigma)$ is convex and continuously differentiable in $\sigma$. 
We stop iteration when $\hat\sigma$ gets converged. Due to the joint convexity of (\ref{eq:main2supp}), the estimates produced from the above iterative algorithm converge to the minimizer of (\ref{eq:main2supp}), with $\widehat\B^n=\widehat\B^n(\hat\sigma\w)$.
In step (\ref{eq:ite1}), the optimization problem is solved by using an ADMM based algorithm described in \citet{li2018integrative}.  %where in each iteration we update $\B$ and $\sigma$ sequentially in the primal step. The algorithm is introduced in detail in section \ref{sec:com}.   

%\subsection{ADMM for Scaled iRRR}

An alternative is to directly apply ADMM to solve (\ref{eq:main2supp}). Let $\A_k\ (k=1,\ldots,K)$ be the surrogate variables of $\B_k$ with the same dimension, we optimize
\begin{align*}
\min_{\A_k,\B_k,\sigma} & \frac{1}{2nq\sigma}\left\|\Y-\sum_{k=1}^K \X_k\B_k \right\|_F^2 + \frac{\sigma}{2} + \lambda \sum_{k=1}^K w_k \|\A_k \|_* \\
\text{s.t.} \ & \A_k = \B_k,\ k=1,\ldots,K.
\end{align*}
Let $\bLambda = (\bLambda_1\trans, \ldots, \bLambda_K\trans)\trans$ be the Lagrange parameter with each $\bLambda_k \in \mathbb{R}^{p_k\times q}$, then the augmented Lagrangian objective function is
\begin{align*}
D(\Y; \A, \B, \sigma, \bLambda) & =\frac{1}{2nq\sigma}\left\|\Y-\sum_{k=1}^K \X_k\B_k\right\|_F^2 + \frac{\sigma}{2} + \lambda \sum_{k=1}^K w_k \|\A_k\|_* \\
& +  \sum_{k=1}^K \left<\bLambda_k, \A_k - \B_k\right>_F + \frac{\rho}{2}\sum_{k=1}^K \left\|\A_k-\B_k\right\|_F^2,
\end{align*}
where $\rho$ is a pre-specified constant to control the step size. Let $\widetilde \A$, $\widetilde \B,\ \tilde\sigma$ and $\widetilde \bLambda$ be the estimates from the last iteration, then in the primal step we first update $(\B,\sigma)$ with the given $\widetilde \A$ and $\widetilde \bLambda$, secondly estimate $\A$ based on the updated $(\B,\sigma)$ and $\widetilde \bLambda$, and finally conduct the dual step. Specifically, to update $(\B,\sigma)$ we first estimate $\B$ %based on $\tilde\sigma$, $\widetilde\A$ and $\widetilde \bLambda$ 
with 
\begin{align}\label{estb}
	\widehat \B = \left( \frac{1}{nq\tilde\sigma} \X\trans\X + \rho \I_p \right)^{-1} \left( \frac{1}{nq\tilde\sigma} \X\trans\Y + \widetilde\bLambda + \rho \widetilde\A \right),
\end{align}
and then update $\sigma$ with
\begin{align}\label{estsigma}
	\hat\sigma = \frac{\|\Y - \X \widehat \B\|_F}{\sqrt{nq}}.
\end{align}
Here we remark that we only update $(\B,\sigma)$ once in each iteration, and it works well in simulation. Since the objective function is separable with respect to each $\A_k$ given $(\widehat\B,\hat\sigma,\widetilde\bLambda)$, we have  
$$
D(\Y, \A_k, \widehat \B, \hat\sigma, \widetilde \bLambda)=\lambda w_k \|\A_k\|_* + \left<\widetilde\bLambda_k,\A_k\right>_F + \frac{\rho}{2}\|\A_k\|_F^2-\rho \left<\A_k,\widehat\B_k\right>_F.
$$
Minimizing the above function with respect to $\A_k$ is equivalent to solving
$$
\min_{\A_k} \frac{1}{2}\|\A_k - (\widehat\B_k - \widetilde\bLambda_k/\rho)\|_F^2 + \frac{\lambda w_k}{\rho}\|\A_k\|_*,
$$
which has an explicit solution \citep{cai2010singular}
\begin{align}
\widehat{\A_k}=\U_k \mathcal{S}(\D_k, w_k \lambda/\rho)\V_k\trans, \label{eq:Astep}
\end{align}
where $\U_k$, $\V_k$ and $\D_k$ come from the singular value decomposition $(\widehat\B_k - \widetilde\bLambda_k/\rho)=\U_k\D_k\V_k\trans$, and $\mathcal{S}(\D_k,w_k \lambda /\rho)=(\D_k - w_k \lambda/\rho)_+$ is the soft-thresholding operator to all the diagonal elements of $\D_k$. 
Finally, based on the updated $\widehat\A_k$ and $\widehat\B_k$, the dual step is
\begin{align}
\widehat \bLambda_k = \widetilde \bLambda_k + \rho \left(\widehat \A_k-\widehat \B_k\right), \ k=1,\ldots,K. \label{eq:Dstep}
\end{align}
For establishing the stopping rule, the primal residual and dual residual are defined as 
\begin{align}\label{eq:residual}
r_{primal}&=\| \widehat\A - \widehat\B\|_F, \nonumber \\
r_{dual}&=\rho \|\widehat \A - \widetilde \A\|_F.
\end{align}
Once both residuals fall below a pre-specified tolerance level, we stop the iteration. In practice, we can gradually increase the step size $\rho$ to accelerate the algorithm \citep{he2000alternating}. The procedure is summarized in Algorithm \ref{alg1}.

\begin{algorithm}[h]
\caption{The ADMM algorithm to solve \eqref{eq:main2supp}.}\label{alg1}
\begin{algorithmic}
\State Parameter: $\lambda$, $\rho$.
\State Initialize $\A$, $\B$, $\sigma$ and the Lagrange multiplier $\bLambda$;
\While {The stopping criterion is not satisfied}
\bi
\item Primal step: 
    \bi
    \item Update $\B_k, \ k=1,\ldots,K$ by \eqref{estb};
    \item Update $\sigma$ by \eqref{estsigma};
    \item Update $\A_k, \ k=1,\ldots,K$ by \eqref{eq:Astep};
    \ei
\item Dual step:
  \bi
  \item Update $\bLambda$ by \eqref{eq:Dstep};
  \ei  
\item Calculate the primal and dual residuals defined in \eqref{eq:residual};
\item (Optional) Increase $\rho$ by a small amount, e.g., $\rho \leftarrow 1.01\rho$.
\ei
\EndWhile
\end{algorithmic}
\end{algorithm}

\subsection{Score Matrix Estimation}\label{web:A2}
 
In order to obtain the score matrix, we need to solve (\ref{inf8supp}) which can be formulated as 
\begin{align}\label{opt2}
\min_{\A_k,\B_k} & \frac{1}{2n}\|\Y-\sum_{k=1}^K \X_k\B_k \|_F^2 + \lambda \sum_{k=1}^K w_k \|\X_k\A_k \|_* \\
\text{s.t.} \ & \X_k\A_k = \X_k \B_k,\ k=1,\ldots,K. \nonumber
\end{align}
Different from the original iRRR optimization problem \eqref{eq:main1}, in (\ref{opt2}) we penalize the nuclear norm of the group effect $\X_k\B_k$ directly. The ADMM algorithm proposed in \citet{li2018integrative} can be applied here with a small modification, i.e., based on $\widehat\B$ and $\widetilde \bLambda$ we need to update $\X_k\A_k$ but not $\A_k$ from 
\begin{align*}
	\min_{\A_k\X_k} \frac{1}{2}\|\X_k\A_k - (\X_k\widehat\B_k -  \widetilde\bLambda_k/\rho)\|_F^2 + \frac{\lambda w_k}{\rho}\|\X_k\A_k\|_*.
\end{align*}
By conducting singular value decomposition to $ \X_k\widehat\B_k - \widetilde\bLambda_k/\rho $ and applying soft thresholding to its singular values with the threshold value $\lambda w_k/\rho$ we can update $\X_k\A_k$. Accordingly, we have 
\begin{align*}
	& \widehat \bLambda_k = \widetilde \bLambda_k + \rho(\widehat{\X_k\A_k} - \X_k\widehat \B_k), \\
	& r_{primal}=\| \widehat{\X\A} - \X\widehat\B\|_F, \\
    & r_{dual}=\rho \sum_{k=1}^K \|\X_k\trans(\widehat{\X\A} - \widetilde{\X\A})\|_F.
\end{align*}
Once both residuals fall below a pre-specified tolerance level we stop the algorithm.

%\section*{Appendix}\label{appendix}
\section{Proof of Theorem \ref{thm:scaleiRRR}}\label{web:B} %\ref{thm:scaleiRRR}}
\begin{proof}[Proof] %\ref{thm:scaleiRRR}] 
% %Recall that we have $(\widehat \B^n,\hat\sigma)$ be the minimizer of (\ref{eq:main2supp}) and $\widehat \B^n=\widehat\B^n(\hat\sigma \w)$. Then w
We follow the proof in \citet{mitra2016benefit}. With $\eta>0$, first define
\begin{align}
\mu(\w,\eta)=\frac{8(1+\eta)(2+\eta)}{\eta^2} \frac{\sum_{k=1}^K  q r_k \lambda^2 w_k^2}{\kappa(\X)},\ \tau_+=\frac{2+\eta}{1+\eta} \mu(\w,\eta), \ \tau_-=\frac{2}{1+\eta}\mu(\w,\eta),	\label{def1}
\end{align}
and an event 
	$$\mathcal{E} =  \cap_{k=1}^K \mathcal{A}_k = \cap_{k=1}^K \left\{\frac{d_1(\X_k\trans\E)}{nq \sigma^*/\sqrt{1+\tau_-}}   \leq \frac{\lambda w_k}{1+\eta}\right\}.$$
Let $\bDelta = \widehat \B^n(t\w) - \B^*$, $\bDelta_k = \widehat \B^n_k(t\w) - \B_k^*$, $\hat\sigma^2(t\w)=\|\Y - \X\widehat \B^n(t\w)\|_F^2/(nq)$ and $t \geq \sigma^*/\sqrt{1+\tau_-}$, then
	\begin{align}
	\sigma^{*2}-\hat\sigma^2(t\w) & = \frac{\|\Y - \X\B^*\|_F^2}{nq} - 	\frac{\|\Y - \X\widehat \B^n(t\w)\|_F^2}{nq} \nonumber \\
	& = \frac{\langle\X\bDelta, 2\E -\X\bDelta\rangle _F}{nq} \nonumber \\
	& = \frac{\langle \X\bDelta, \Y + \E -\X\widehat \B^n(t\w)\rangle _F}{nq} \nonumber \\
	& = \frac{\langle \X\bDelta, \E\rangle _F}{nq} + \frac{\langle \X\bDelta, \Y - \X\widehat \B^n(t\w)\rangle _F}{nq}  . \label{eq:proof1}
	\end{align}
We first deal with the first term on the right hand side of (\ref{eq:proof1}) and get
\begin{align}
\frac{|\langle \X\bDelta, \E\rangle _F|}{nq} & %= \frac{|\langle \sum_{k=1}^K\X_k\bDelta_k, \E\rangle _F|}{nq} \nonumber \\
%& \leq \frac{\sum_{k=1}^K|\langle \X_k\bDelta_k, \E\rangle _F|}{nq}	 \nonumber \\
 \leq \sum_{k=1}^K \frac{d_1(\X_k\trans\E)\|\bDelta_k\|_*}{nq} \nonumber \\
& \leq \frac{\lambda t}{1+\eta}\sum_{k=1}^K w_k \|\bDelta_k\|_*. \label{eq:proof2}
\end{align}
The last inequality is built on the event $\mathcal{E}$ with $t \geq \sigma^*/\sqrt{1+\tau_-}$. Next we deal with the second term on the right hand side of (\ref{eq:proof1}) and obtain 
\begin{align*}
\frac{|\langle \X\bDelta, \Y - \X\widehat \B^n(t\w)\rangle _F|}{nq} %& = \frac{1}{nq} \left|\left\langle \sum_{k=1}^K \X_k \bDelta_k, \Y -\X\widehat \B^n(t\w) \right\rangle _F\right| \\
%& \leq \frac{1}{nq} \sum_{k=1}^K|\langle \X_k \bDelta_k, \Y -\X\widehat \B^n(t\w) \rangle _F| \\
%& 
\leq \frac{1}{nq} \sum_{k=1}^K \|\bDelta_k\|_* d_1(\X_k\trans\{\Y -\X\widehat \B^n(t\w)\}).
\end{align*}
%After deriving the subgradient of $\L_{\w}(\B)$, 
In order to bound $d_1(\X_k\trans\{\Y -\X\widehat \B^n(t\w)\})$, recall that $\widehat \B^n(t\w)$ is a minimizer of $\L_{t\w}(\B)$ if and only if there exists a diagonal matrix $\J_k$ with $d_1(\J_k) \leq 1$ such that 
$$
\X_k\trans\{\Y-\X\widehat \B^n(t\w)\}=\lambda t n q w_k \U_k \J_k \V_k\trans,\ k=1,\ldots,K,
$$ 
where $\widehat \B^n_k(t\w)=\U_k\D_k\V_k\trans$ is the singular value decomposition \citep{watson1992characterization}. Thus for each $k$, we have
$$
d_1(\X_k\trans\{\Y-\X\widehat \B(t\w)\}) \leq \lambda tnqw_k,
$$
and 
\begin{align}
\frac{|\langle\X\bDelta, \Y - \X\widehat \B^n(t\w)\rangle _F|}{nq} \leq t\lambda\sum_{k=1}^K w_k \|\bDelta_k\|_*. \label{eq:proof3}
\end{align}
Then from 
$$
\frac{\langle\X\bDelta, 2\E -\X\bDelta\rangle _F}{nq} \leq \frac{\langle\X\bDelta, 2\E\rangle _F}{nq}
$$
and inequality (\ref{eq:proof2})
we have
\begin{align}
\sigma^{*2}-\hat \sigma^2(t\w) \leq \frac{|\langle\X\bDelta, 2\E\rangle_F|}{nq}\leq \frac{2t\lambda}{1+\eta}\sum_{k=1}^K w_k \|\bDelta_k\|_*,	
\end{align}
and from (\ref{eq:proof1}), (\ref{eq:proof2}) and (\ref{eq:proof3}), we have
\begin{align}
\sigma^{*2}-\hat \sigma^2(t\w) & \geq -\frac{|\langle\X\bDelta, \Y - \X\widehat \B(t\w)\rangle _F|}{nq} - \frac{|\langle\X\bDelta, \E\rangle _F|}{nq} \nonumber \\
& \geq -\frac{t\lambda(2+\eta)}{1+\eta}\sum_{k=1}^K w_k \|\bDelta_k\|_*.
\end{align}
Therefore, we have
\begin{align}
	-\frac{t\lambda(2+\eta)}{1+\eta}\sum_{k=1}^K w_k \|\bDelta_k\|_*\leq \sigma^{*2}-\hat \sigma^2(t\w) \leq \frac{2t\lambda}{1+\eta}\sum_{k=1}^K w_k \|\bDelta_k\|_*.\label{bound1}
\end{align}
Next, we derive the rate of $\sum_{k=1}^Kw_k\|\bDelta_k\|_*$ by analyzing 
\begin{align}
\frac{\widehat \B^n(t\w)}{t} = \arg\min_{\B \in \mathbb{R}^{p \times q}} \left\{\frac{1}{2nq}\|\Y/t -\X\B\|_F^2 + \lambda\sum_{k=1}^K w_k \|\B_k\|_* \right\}. \label{bovert}
\end{align}
%which estimates $\B^*/t$ from data $(\Y/t,\X)$.	
Since $t \geq \sigma^*/\sqrt{1+\tau_-}$, event $\mathcal{E}$ leads to  
$$
\cap_{k=1}^K \left\{\frac{d_1(\X_k\trans\E/t)}{nq} \leq \frac{\lambda w_k}{1+\eta} \right\},
$$	
which facilitates the application of Theorem 2 in \citet{li2018integrative} to \eqref{bovert} and we obtain 
\begin{align}
	t^{-1}\lambda\sum_{k=1}^K w_k \|\bDelta_k\|_* = \lambda\sum_{k=1}^K w_k \|\widehat \B_k^n (t\w)/t - \B^*_k/t\|_* \leq \frac{8(1+\eta)(2+\eta)}{\eta^2} \frac{\sum_{k=1}^K  r_k q \lambda^2 w_k^2}{\kappa(\X)}=\mu(\w,\eta). \nonumber
\end{align}
It follows that
\begin{align}
t\lambda\sum_{k=1}^K w_k \|\bDelta_k\|_* \leq t^2 \mu(\w,\eta), \label{eq:proof4}
\end{align}
which together with \eqref{bound1} leads to
\begin{align*}
	-\frac{2+\eta}{1+\eta} t^2 \mu(\w,\eta) \leq \sigma^{*2}-\hat\sigma^2(t\w) \leq \frac{2}{1+\eta} t^2 \mu(\w,\eta).
\end{align*}
Recall the definition of $\tau_+$ and $\tau_-$ in (\ref{def1}), we have
\begin{align}
	-\tau_+ t^2 \leq \sigma^{*2}-\hat \sigma^2(t\w)\leq \tau_-t^2.\label{proof5}
\end{align}
The second inequality in (\ref{proof5}) with $t=\sigma^*/\sqrt{1+\tau_-}$ leads to $t^2-\hat\sigma^2(t\w) \leq t^2 - \sigma^{*2} + \tau_-t^2 =0 $, which indicates $\hat \sigma(t\w) \geq t = \sigma^*/\sqrt{1+\tau_-}$. Assume $\sigma^*/\sqrt{1-\tau_+} \geq \sigma^*/\sqrt{1+\tau_-}$, then the first inequality of (\ref{proof5}) with $t=\sigma^*/\sqrt{1-\tau_+}$ implies $t^2-\hat\sigma^2(t\w) \geq t^2 - \sigma^{*2} - \tau_+t^2 =0 $, i.e., $\hat \sigma(t\w) \leq t = \sigma^*/\sqrt{1-\tau_+}$. Due to the joint convexity of the scaled iRRR framework (\ref{eq:main2supp}), we have $\hat\sigma(t\w)=\|\Y - \X\widehat \B^n(t\w)\|_F/\sqrt{nq}$ converges to $\hat\sigma$ which is the minimizer of (\ref{eq:main2supp}) and consequently we have
\begin{align*}
	\frac{\sigma^*}{\sqrt{1+\tau_-}} \leq \hat \sigma \leq \frac{\sigma^*}{\sqrt{1-\tau_+}},
\end{align*}
which is followed by 
\begin{align*}
	\left|\frac{\hat \sigma}{\sigma^*}-1\right|=o_p(\mu(\w,\eta)).
\end{align*}
If we have $\sqrt{nq}\mu(\w,\eta) \rightarrow 0$, then
\begin{align*}
	\left|\frac{\hat \sigma}{\sigma^*}-1\right|=o_p((nq)^{-1/2}).
\end{align*}
Moreover, if $vec(\E)\sim \mathcal{N}_{nq}(0, \sigma^2\I_{nq})$, we have 
$
\sigma^*/\sigma \sim \chi_{nq}/\sqrt{nq}.
$ 
Then by central limit theorem we get
$$
\sqrt{nq} \left(\frac{\sigma^*}{\sigma}-1\right) \rightarrow \mathcal{N}\left(0,\frac{1}{2}\right).
$$
Consequently, we can prove
\begin{align}
\sqrt{nq} \left(\frac{\hat\sigma}{\sigma}-1\right) \rightarrow \mathcal{N}\left(0,\frac{1}{2}\right)\label{asympnormalsupp}.
\end{align}
Next we derive the estimation error bound for $\widehat \B^n(\hat \sigma \w)$ (i.e., $\widehat \B^n$) under the framework \eqref{bovert}. Since $\hat \sigma \geq \sigma^*/\sqrt{1+\tau_-}$, the estimation error bounds are 
\begin{align*}
	\| \widehat {\B}^n - \B^* \|_F^2  \preceq \frac{  \sigma^{*2}\sum_{k=1}^K r_k q^2 \lambda^2 w_k^2 }{(1-\tau_+)\kappa^2(\X)}, \\ 
	\sum_{k=1}^K \lambda w_k \|\widehat {\B}_k^n - \B_k^*\|_*  \preceq \frac{\sigma^{*}\sum_{k=1}^K r_k  q \lambda^2 w_k^2 }{\sqrt{1-\tau_+}\kappa(\X)}
\end{align*}
by applying Theorem 2 in \citet{li2018integrative} and the fact that $\hat \sigma \leq \sigma^*/\sqrt{1-\tau_+}$.
Finally, we need to prove $\P(\mathcal{E})> 1-\epsilon$ with some $0<\epsilon<1$. Note that follow the same reasoning as the proof of Theorem 6 in \citet{mitra2016benefit} with the assumption $vec(\E)\sim \mathcal{N}_{nq}(0, \sigma^2\I_{nq})$, it can be verified that if we let $w_{k} =d_1(\X_k)\left\{\sqrt{p_k/n} + \sqrt{2\log(K/\epsilon)/(nq)}\right\}/\sqrt{nq}$ with a properly selected $\lambda$, we have $\P(\mathcal{E})> 1-\epsilon$. This completes the proof.

\end{proof}

\section{A Brief Overview of High-Dimensional Inference Procedures and the LDPE Approach}\label{web:C}

%\subsection{A Brief Overview of High-Dimensional Inference Procedures}

Researches on statistical inference for regularized estimators emerged in recent years as the prevailing of high-dimensional statistics. Regularized estimation methods are commonly used in high dimensional linear regression problems, e.g., lasso \citep{tib1996}, elastic net \citep{zou2005regularization}, and group lasso \citep{yuan2006model}. %(Change to ``lasso''. Also, for this literature review, it may be better to first give a general picture: what are the main approaches of making inference? Then provide some key references and discussions for each.)
However, due to the regularization, the resulting estimator is often biased and not in an explicit form, making its sampling distribution complicated and even intractable. In order to account for uncertainty in estimation and assess the selected model, several methods have been proposed for assigning p-values and constructing confidence intervals for a single or a group of coefficients. See, e.g., \citet{knight2000asymptotics, wasserman2009high, meinshausen2009p, chatterjee2013rates, ning2017general, shi2019linear}. % For example, \citet{knight2000asymptotics} investigated the limiting distribution of lasso estimator. \citet{wasserman2009high} considered a random-splitting method. % where half of the data are used for variable selection and then p-values are produced by applying the selected model to another half of the data. %And in order to solve the sensitivity of the random-splitting method to random splits, \citet{meinshausen2009p} perfected this method with a multi sample-splitting procedure. 
 %\citet{ning2017general} and \citet{shi2019linear} proposed several approaches by extending the classical likelihood ratio test, Wald test and score test to the high-dimensional situation. %For example,\citet{ning2017general} extended the score test to a high dimensional case for generic penalized M-estimators with a newly proposed decorrelated score function. To test linear hypotheses in the high dimensional generalized linear models, \citet{shi2019linear} proposed several test methods based on a constrained partial regularized estimator. 
  %There are also some other methods, e.g., \citet{chatterjee2013rates} and \citet{liu2013asymptotic} considered residual bootstrap method to conduct valid statistical inference, \citet{lockhart2014significance} proposed a covariance test statistic for testing of lasso, %\citet{belloni2014uniform}
  % \citet{meinshausen2010stability} and \citet{shah2013variable} combined subsampling and selection algorithm to control false discovery rate. 
  %Moreover, for dealing with the situation where strong correlation is among covariates and single coefficient inference is impossible, \citet{meinshausen2015group} proposed a group-bound method to detect effects from a group of variables without regularity conditions on the design. 
 One popular class of method utilizes the projection and bias-correction technique, where a de-biasing procedure is first applied to the regularized estimator and then the asymptotic distribution is derived for the resulting estimator. For example, \cite{buhlmann2013statistical} applied bias correction to a Ridge estimator and derived an inference procedure, \cite{zhang2014confidence} and \citet{javanmard2014confidence} considered the lasso estimator. \cite{shi2016regression} generalized the procedure of \cite{javanmard2014confidence} to make inference for lasso estimator obtained under multiple linear constraints on coefficients. To facilitate chi-square type hypothesis testing for a possibly large group of coefficients without inflating the required sample size due to group size,
  \cite{mitra2016benefit} generalized the idea of the low-dimensional projecting estimator \citep[LDPE]{zhang2014confidence} to correct the bias of a scaled group lasso estimator. %For more inference methods see \citet{dezeure2015high}, where a comprehensive simulation study is also included to conduct comparison in terms of empirical performance of several different inference methods.
Although the above methods are effective under various model settings, to the best of our knowledge, so far there is not much work focus on inference in high-dimensional multivariate regression, especially for rank restricted models, which motivates the derivation of the inference method considered in this paper. %  in this paper. %The proposed method will be built upon the LDPE approach proposed by \citet{zhang2014confidence}.

It is worthwhile to dive deeper into the LDPE approach proposed by \citet{zhang2014confidence}, as our proposed method will be built upon it. The illustration proceeds under a multiple linear regression model $\y=\x_1\beta_1 + \ldots + \x_p\beta_p + \epsilon$, where $\y\in \mathbb{R}^n$ is the response vector and $\x_j\in \mathbb{R}^n$ is a vector consisting of observations of the $j$-th predictor. Suppose we are interested in the effect of predictor $\x_j\ (1\leq j \leq p)$ on the response. The initial estimator $\hat\beta^n_j$ can be obtained by lasso method. As we mentioned before, lasso estimator is biased due to the regularization on coefficients. The effect of $\x_j$ on response cannot be fully represented by $\hat \beta^n_j$, hence a properly selected score vector is used to recover the part of information that is lost in regularization. The resulting LDPE $\hat \beta_j$ can be written as 
\begin{align}\label{de-bias}
	\hat \beta_j = \hat \beta_j^n + \frac{\z_j\trans(\y-\sum_{l=1}^p \x_l \hat \beta_l^n)}{\z_j\trans\x_j},
\end{align}
where the score vector $\z_j$ has the same dimension as $\x_j$ and only depends on the design matrix $\X=(\x_1,\ldots,\x_p)$. The score vector $\z_j$ serves as a tool to extract the information that is only related to $\x_j$ from the residual, then to correct the bias this part of effect is added back to $\hat \beta^n_j$ after standardization. 

The classical scenario with $n > p$ can help us understand the mechanism of the above procedure better. When $n > p$, $\z_j$ can be set as $\x_j^{\perp}$, the projection of $\x_j$ onto the orthogonal complement of the column space of $\X_{-j}$ (the design matrix with the $j$-th column deleted). This choice of $\z_j$ can be regarded as the information only carried by the $j$-th predictor and satisfies $\z_j\trans\X_{-j}=\0$. Then whatever the initial estimator is, the resulting de-biased estimator is the least square estimator which is unbiased. However, in order to satisfy $\z_j\trans\X_{-j}=\0$ when $p < n$, $\z_j$ needs to be a zero vector which consequently makes \eqref{de-bias} ineffective. Therefore, in order to apply the de-biasing procedure \eqref{de-bias} in the high-dimensional scenario, we have to relax the requirement $\z_j\trans\X_{-j}=\0$, i.e., to approximate $\x_j^{\perp}$ to control the bias caused by $\z_j\trans\X_{-j}\neq \0$ to be under a tolerable level. For example, the score vector is obtained by applying lasso to the regression of $\x_j$ on $\X_{-j}$ in \citet{zhang2014confidence}, while in \citet{javanmard2014confidence} the score vector is estimated from an optimization program that minimizes the variance of the de-biased estimator while control its bias.

\section{Derivation of The Inference Procedure}\label{web:D}

In this section, we introduce the main steps of establishing our proposed method follow \citet{mitra2016benefit}, which include (1) the exploitation of LDPE to correct the bias of the scaled iRRR estimator, (2) the construction of a $\chi^2$-type test statistic based on the de-biased estimator, and (3) the estimation of the required score matrix and the derivation of the theoretical guarantee of the reliability of the test. The condition $\mbox{rank}(\S_k\trans\X_k)=\mbox{rank}(\X_k)$ is required to guarantee the effectiveness of de-biasing, under which the role of $\S_k$ in the de-biasing procedure can be totally replaced by $\P_k$. %Thus in the following we assume $\mbox{rank}(\S_k\trans\X_k)=\mbox{rank}(\X_k)$ and use $\P_k$ as the score matrix. 
%{\color{red} (Add a sentence of short summary of the major steps?)}
%where $w_k$'s are some pre-specified weights to adjust for the strength of regularization on each view and $\lambda$ is a uniform tuning parameter. %Due to the using of a composite low-rank penalty term $\sum_{k=1}^{K} w_k \|\B_k\|_*$, low-rank structures are imposed on each coefficient sub-matrix to induce sparsity in the model, and it also nicely bridges lasso, group lasso and nuclear norm penalized regression methods.

First, we provide the de-biased scaled iRRR estimator based on the notations defined in the main paper. 
%Let $\S_k\in \mathbb{R}^{n \times p_k}$ be the score matrix of $\X_k$ that only depends on $\X$. {\color{red}This is a critical tool used in LDPE to correct the bias caused by regularization and will be specified later.} Write $\Q_k$ and $\P_{0,k}$ be the orthogonal projection matrices onto the column spaces of $\X_k$ ($\mathbb{C}(\X_k)$) and $\S_k$ ($\mathbb{C}(\S_k)$), respectively, and let $\P_k$ be the projection matrix of $\mathbb{C}(\P_{0,k}\Q_k)$. Hereafter, we assume $\mbox{rank}(\S_k\trans\X_k)=\mbox{rank}(\X_k)$, {\color{red}which guarantees the effectiveness of the de-biasing procedure.}
With the scaled iRRR estimator $\widehat \B^n=(\widehat \B_1\ntrans,\ldots,\widehat\B_K\ntrans)\trans$ from (\ref{eq:main2supp}), the de-biased estimator of $\B_k$ is
\begin{align}
	\widehat \B_k=\widehat \B_k^n + (\S_k'\X_k)^+\S_k\trans(\Y-\X\widehat\B^n), %\label{ldpe1}
\end{align}   
where $(\S_k'\X_k)^+$ is the Moore-Penrose inverse of $\S_k'\X_k$. For the group effect $\X_k\B_k$, the related de-biased estimator is 
\begin{align}
	\X_k\widehat \B_k & = \X_k\widehat \B_k^n + (\P_k\Q_k)^+\P_k(\Y-\X\widehat\B^n). %\label{ldpe2}
\end{align}

Next, based on the de-biased estimator, we introduce a test statistic and derive its asymptotic distribution under the null. Note that,
if $\mbox{rank}(\X_k)=p_k$ we have
\begin{align}
	(\P_k\X_k)(\widehat \B_k - \B_k^*)=\P_k\E - \mbox{Rem}_k \label{inf1}
\end{align}
with
\begin{align}%\label{remk}
	\mbox{Rem}_{k}=\P_k\sum_{j \neq k} (\X_j\widehat \B_j^n-\X_j\B_j^*),
\end{align}
and if $\mbox{rank}(\X_k)< p_k$, we can only make inference on $\X_k\B_k^*$ with
$$(\P_k\Q_k)(\X_k\widehat \B_k - \X_k\B_k^*)=\P_k\E - \mbox{Rem}_k.$$
The effect of de-biasing in $\widehat \B_k$ and $\X_k\widehat\B_k$ is controlled by the approximation of $\S_k$ to $\X_k^{\perp}$ and the distance between $\widehat\B^n$ and $\B^*$, where $\X_k^{\perp}$ is the best score matrix only available in the `low-dimensional' scenario and is defined as the projection of $\X_k$ onto the orthogonal complement of the column space spanned by $(\X_1,\ldots,\X_{k-1},\X_{k+1},\ldots,\X_K)$. These two factors can be jointly measured by $\mbox{Rem}_{k}$. Therefore, once the magnitude of $\mbox{Rem}_{k}$ is ignorable in the sense that 
\begin{align}
 	\sqrt{qr_k'}|\sigma/\hat\sigma-1|+\|\mbox{Rem}_k/\sigma\|_F=o_p(1), \label{inf04}
 \end{align}
% where $\hat\sigma$ is a consistent noise level estimator obtained from (\ref{eq:main2supp}) and $r_k'=\mbox{rank}(\P_k)=\mbox{rank}(\X_k)$, then we have
we have the approximation $\|\P_k\E-\mbox{Rem}_k\|_F^2/\hat\sigma^2 \rightarrow \|\P_k\E/\sigma\|_F^2$,
which together with the normal assumption on the random error matrix %that  
% \begin{align*}
% 	vec(\E) \sim \mathcal{N}_{nq}(\0, \sigma^2 \I_{n q}),
% \end{align*}
 implies $\|\P_k\E-\mbox{Rem}_k\|_F^2/\hat\sigma^2 \rightarrow \chi^2_{r_k'q}$ where $r_k'=\mbox{rank}(\P_k)=\mbox{rank}(\X_k)$.
%the pivotal statistic $\|\P_k\E-\mbox{Rem}_k\|_F^2/\hat\sigma^2$ asymptotically follows $\|\P_k\E/\sigma\|^2_F \sim \chi^2_{r_k'q}$.
%And if (\ref{inf4}) is satisfied, then we can also use $\chi^2_{r_k'q}$ as the null distribution since $\|\P_k\E\|_F^2/\sigma^2 \sim \chi^2_{r_k'q}$.
If in the true model $\B^*_k=\0$ or $\X_k\B^*_k=\0$, then since $\Y=\X_{-k}\B_{-k}^*+\E$ we have 
\begin{align}
 \P_k\E - \mbox{Rem}_k=\P_k(\Y-\X_{-k}\widehat \B_{-k}^n). \label{inf2}
\end{align}
Therefore, the test statistic is
\begin{align}
	T_k=\frac{1}{\hat \sigma^2}\left\|\P_k(\Y-\sum_{j \neq k}\X_{j}\widehat \B_{j}^n)\right\|_F^2 \overset{H_0}{\sim} \chi^2_{r_k'q}
\end{align}
asymptotically. We shall note that if $\mbox{rank}(\X_k)<p_k$, $\B_k^*$ is not identifiable, the method is only applicable to test $\mbox{H}_0:\X_k\B^*_k=\0\ \mbox{vs.}\ \mbox{H}_1: \X_k\B^*_k \neq \0$ and when $\mbox{rank}(\X_k)=p_k$, the method is also applicable to test $\mbox{H}_0:\B^*_k=\0\ \mbox{vs.}\ \mbox{H}_1: \B^*_k \neq \0$. %The details of the derivation of the test is provided in Appendix. 

In order to implement and validate this test procedure, in addition to the scaled iRRR estimator $(\widehat\B^n,\hat\sigma)$, we also need to find $\P_k$ and verify condition \eqref{inf04}. One key ingredient to verify (\ref{inf04}) is to make $\|\mbox{Rem}_k/\sigma\|_F=o_p(1)$. 
 Recall the form of $\mbox{Rem}_k$, we have
 \begin{align}
 	\frac{\|\mbox{Rem}_k\|_F}{\sigma\sqrt{nq}} & \leq \frac{\sum_{j\neq k} \|\P_k \X_j (\widehat \B_j^n -\B^*_j)\|_F}{\sigma\sqrt{nq}}   \nonumber \\ 
 	& \leq \sum_{j\neq k} \frac{d_1(\P_k\Q_j)}{w_{*,j}\sigma\sqrt{nq}}w_{*,j}\|\X_j\widehat \B_j^n - \X_j\B^*_j\|_F \nonumber \\
 	& \leq \max_{j\neq k} \frac{d_1(\P_k\Q_j)}{w_{*,j}} \sum_{j=1}^K \frac{w_{*,j}}{\sigma\sqrt{nq}} \|\X_j\widehat \B^n_j - \X_j\B^*_j\|_F \nonumber \\
	& =O_p(q \sum_{j=1}^K r_j w_j^2) \max_{j\neq k} \frac{d_1(\P_k\Q_j)}{w_{*,j}}, \nonumber
\end{align}
which leads to 
\begin{align}\label{proofthm2}
	\frac{\|\mbox{Rem}_k\|_F}{\sigma} = O_p\left(\sum_{k=1}^K \frac{r_k \{ p_k q + 2\log(K/\epsilon) \}}{\sqrt{nq}}\right) \eta_k
\end{align}
where $\eta_k=\max_{j\neq k} d_1(\P_k\Q_j)/w_{*,j}$ is dominated by $d_1(\P_k\Q_j)$. Thus, an ideal $\P_k$ needs to minimize the variance of the resulting de-biased estimator while control the magnitude of $d_1(\P_k\Q_j)$. 
\citet{mitra2016benefit} derived the following optimization framework
\begin{align}
	\P_k=\arg\min_{\P} \{d_1(\P(\I-\Q_k)): \P=\P^2=\P\trans,\ d_1(\P\Q_j) \leq w_j', \forall j\neq k  \} \label{inf7}
\end{align}
to solve out $\P_k$. In (\ref{inf7}), $w_j'$ is an upper bound of $d_1(\P\Q_j)$ and $d_1(\P_k(\I-\Q_k))$ measures the distance between the subspaces spanned by $\P_k$ and $\I-\Q_k$, which may inflate the variance of the de-biased estimator. The feasibility of (\ref{inf7}) with a given $w_j'$ has been verified for random designs with sub-Gaussian rows, refer to Theorem 4 and Lemma 1 in \citet{mitra2016benefit} for details. Since (\ref{inf7}) has not been solved yet, in practice, $\P_k$ can be estimated from a penalized multivariate regression (\ref{inf8supp}). Then with the conditions in Theorem \ref{thm2} of the main paper, we can verify (\ref{inf04}) thus validate the inference procedure.

\section{Proof of Theorem \ref{thm2}}\label{web:E}%\ref{thm2}}
\begin{proof}[Proof]%\ref{thm2}] 
First we get the rate of $\|\mbox{Rem}_k\|_F/\sigma$. From the KKT condition of (\ref{inf8supp}), we have $d_1(\Q_j \S_k/\sqrt{n} ) \leq \xi w_j^{''}$, which implies $d_1(\Q_j\P_k/\sqrt{n})d_{min}(\S_k) \leq \xi w_j^{''}$. If we let $w_j^{''}=w_{*,j}$, then together with (\ref{proofthm2}) and the condition
$$
\sum_{j=1}^K \frac{r_j \{ p_j q + 2\log(K/\epsilon) \}}{\sqrt{nq}}\left\{\xi d_{min}(\S_k/\sqrt{n})^{-1}  \right\} \rightarrow 0
$$
we have $\|\mbox{Rem}_k\|_F/\sigma=o_p(1)$. Then we consider the rate of $|1-\sigma/\hat\sigma|$. From (\ref{asympnormalsupp}) we %can obtain the asymptotic performance of $\sigma/\hat\sigma$ and 
can get 
\begin{align*}
	\left| 1-\frac{\sigma}{\hat\sigma} \right| = O_p \left( \frac{1}{\sqrt{n q}}\right),
\end{align*}
which together with $r_k'/n \rightarrow 0$ implies 
\begin{align*}
	\left| 1-\frac{\sigma}{\hat\sigma} \right| = o_p\left( \frac{1}{\sqrt{r_k' q}}\right).
\end{align*}
Combine these two results we complete the verification of (\ref{inf04}).  
\end{proof}

\section{Simulation with Real Compositional Data}\label{web:F}
 
%The most representative sample of the preterm infant dataset will be the one directly resampled from the collected microbiome compositional observations with replacement. 
%In this section, a simulation study is conducted based on the resampled real compositional data. 
The data collected from the preterm infant study have the following structure, $n=38$, $p=62$, $q=11$, $K=11$ and the group size is $(p_1,\ldots,p_K)=(3,2,3,4,7,15,2,9,3,2,12)$. From all the 11 groups, we select the fifth, the sixth and the eighth group to be predictive to the response with $r^*_5=r^*_6=r^*_8=2$, and all the remaining groups have no prediction contribution, i.e., $r^*_k=0,\ k \notin \{5,6,8\}$. The results are displayed in Table \ref{comp_simu2}. %Similar to the phenomenon observed in Table \ref{comp_simu1}, 
In general, when the signal is weak, the false positive rates for testing the irrelevant groups are around 0.05 while the power of the test is small. If we increase the SNR, the power increases but the false positive rate will also become larger. As expected, the magnitude of $d_1(\P_k(\I_n-\Q_k))$ has an effect on the performance of the test. Specifically, group 1, 2, 3, 7, 9, 10 have relatively small $d_1(\P_k(\I_n-\Q_k))$ values, and group 4, 5, 6, 8, 11 have $d_1(\P_k(\I_n-\Q_k))$ values close to 1. The magnitude of inflation of the false positive rate for the groups with smaller $d_1(\P_k(\I_n-\Q_k))$ values is smaller than the one for groups with larger $d_1(\P_k(\I_n-\Q_k))$ values. As the preterm infant dataset has a weak signal strength, the simulation results here indicate that we may get relatively conservative but reliable inference results from the application. Moreover, from the comparison between multivariate inference results and univariate inference results, we can observe an improvement in the power of the test by applying a multivariate inference procedure. 
%Group 4 and group 11 are two irrelevant groups with high $d_1(\P_k(\I_n-\Q_k))$ values, and the false positive rates of the corresponding tests become large when the SNR increases. For the irrelevant groups with smaller $d_1(\P_k(\I_n-\Q_k))$ values, the false positive rates of the related tests are smaller. As for the three predictive groups, all of them have large $d_1(\P_k(\I_n-\Q_k))$ values and the related power are relatively low when the signal is weak.

 \begin{table}[ht]
\centering
\caption{Simulation results based on the resampled real microbiome compositional data across 100 replications. The performance of noise level estimation is displayed in terms of the mean ($\times 10$) and standard error ($\times 10$, in parenthesis) of $\hat\sigma/\sigma-1$ and $|\hat\sigma/\sigma-1|$, respectively. Each group is denoted as ``G'' followed by its group number.}\label{comp_simu2}
\resizebox{\linewidth}{!}{%
\begin{tabular}{lccrrrrrrrrrrr}
\hline
\multirow{2}{*}{SNR} & \multirow{2}{*}{$\hat \sigma/\sigma-1$} & \multirow{2}{*}{$|\hat \sigma/\sigma-1|$} & G1 & G2 & G3 & G4 & G5 & G6 & G7 & G8 & G9 & G10 & G11\\
 &   &   &  FP &  FP & FP & FP  &  TP  & TP & FP  & TP & FP & FP & FP\\
\hline
\multicolumn{14}{c}{Multivariate Inference}\\
\hline
1 & 0.61 (0.42) & 0.64 (0.36) & 0.06 & 0.02 & 0.03 & 0.07 & 0.28 & 0.20 & 0.05 & 0.13 & 0.02 & 0.06 & 0.08 \\ 
2 & 1.80 (0.99) & 1.84 (0.93) & 0.08 & 0.10 & 0.13 & 0.11 & 0.80 & 0.83 & 0.07 & 0.50 & 0.03 & 0.10 & 0.34 \\ 
4 & 4.07 (2.08) & 4.09 (2.04) & 0.22 & 0.16 & 0.29 & 0.28 & 0.98 & 0.99 & 0.17 & 0.80 & 0.10 & 0.24 & 0.69 \\ 
\hline
\multicolumn{14}{c}{Univariate Inference}\\
\hline
1 & & & 0.05 & 0.06 & 0.13 & 0.03 & 0.20 & 0.03 & 0.10 & 0.03 & 0.08 & 0.07 & 0.00 \\ 
2 & & & 0.09 & 0.07 & 0.18 & 0.05 & 0.76 & 0.42 & 0.11 & 0.10 & 0.06 & 0.12 & 0.03 \\ 
4 & & & 0.19 & 0.10 & 0.25 & 0.06 & 0.99 & 0.94 & 0.13 & 0.35 & 0.11 & 0.22 & 0.07 \\ \hline
\end{tabular}}
\end{table}

\section{Additional Application Results}\label{web:G}

Table \ref{covariates} lists the estimated coefficients of the control variables from the overall model. In terms of the sub-scale stress/abstinence, the signs of the estimated coefficients are the same as the results from \cite{sun2018log}. Stress/abstinence is the amount of stress and abstinence signs observed in the neurodevelopmental examination procedure \citep{lester2004neonatal}, and a lower value indicates a better neurodevelopment situation. Based on the fitting results, female infants generally perform better in the neurodevelopment examination than male infants. Vaginal delivery and a higher percentage of feeding with mother's breast milk also benefit the neurodevelopment of preterm infants. Moreover, the estimated coefficient of birth weight is -0.029 after multiplying by 1000, which indicates that infants with larger birth weights are more likely to have a better neurological development. SNAPPE-II is one kind of illness severity score, and a higher SNAPPE-II score is often observed among expired infants \citep{harsha2015snappe}. Thus it is reasonable to observe that the SNAPPE-II score is positively related to the stress/abstinence score. As for the PROM, it is a major cause of premature birth and could be very dangerous to both mother and infant. The method provides a positive coefficient estimate of PROM, which matches well with the intuition that a pregnant who did not experience PROM is more likely to give birth to a healthier baby. 
The effects of control variables on other sub-scale scores of NNNS can be similarly interpreted based on the estimated coefficients.
\begin{table}[htp]
\centering
\caption{Estimated coefficients of control variables from the overall model (coefficient of birth weight is multiplied by 1000 and all other coefficients are multiplied by 10).}\label{covariates}
\resizebox{\linewidth}{!}{%
\begin{tabular}{lrrrrrrr}
  \hline
 & Intercept & MBM & Female & Vaginal & PROM & SNAPPE-II & Birth weight \\ 
  \hline
Habituation & 62.373 & 12.694 & 2.394 & 3.317 & -3.020 & 0.069 & -0.517 \\ 
  Attention & 59.367 & -8.039 & 1.308 & 10.392 & 5.081 & -0.032 & -1.103 \\ 
  Handling & 4.908 & -0.139 & -0.739 & -1.508 & -0.441 & 0.015 & 0.132 \\ 
  Qmovement & 39.393 & 4.187 & -3.711 & 3.535 & 2.411 & 0.011 & -0.067 \\ 
  Regulation & 59.774 & 0.448 & -2.745 & 4.578 & 3.821 & -0.256 & -0.665 \\ 
  Nonoptref & 34.916 & 8.465 & -4.702 & -7.855 & -4.039 & 0.264 & 0.998 \\ 
  Stress & 1.901 & -0.032 & -0.088 & -0.265 & 0.029 & 0.015 & -0.029 \\ 
  Arousal & 33.183 & -3.236 & 3.362 & 2.113 & -5.717 & -0.081 & 0.189 \\ 
  Hypotonicity & 1.409 & 8.447 & 2.459 & -2.187 & -4.387 & -0.003 & -0.005 \\ 
  Excitability & 3.088 & -13.781 & 7.811 & -3.680 & -9.722 & 0.179 & 2.397 \\ 
  Lethargy & 7.598 & 34.721 & -0.171 & -23.384 & -0.771 & 0.233 & 2.549 \\ \hline
\end{tabular}}
\end{table}

\clearpage
%%%%%%%%%%%%%%%%%%%%%%%%%%%%%%%%%%%%%%%%%%%%%%%%%%%%
\bibliographystyle{chicago}
%\bibliography{compositional-bibtex,compRRR}

%\newpage
%\appendix

\end{document}